\documentclass[sigconf]{acmart}

\usepackage{paralist}
\usepackage{fancyvrb}
\usepackage{hyperref}
\newcommand{\webref}[2]{\href{#1}{#2}}

\usepackage{xcolor}
\acmConference[BDA '20]{36\`eme Conf\'erence sur la Gestion de Donn\'ees}{October 2020}{Paris}

\usepackage{tikz}
\usepackage{ulem}
\newcommand*\circled[1]{\tikz[baseline=(char.base)]{
            \node[shape=circle,draw,inner sep=0.5pt] (char) {#1};}}

\newtheorem{remark}{Remark}

\newtheorem{theorem}{Theorem}
\newtheorem{example}{Example}
  {\small \em \tt ccc\begin{table} aaa bbb}%
  {\end{table} ddd}
  
\newcommand{\Tr}[1]{\BeginTr{#1}\EndTr}  
\newcommand{\BeginTr}{\langle}  
\newcommand{\EndTr}{\rangle}  

\newcommand{\D}{\,\text{-}\,}

\newcommand{\json}{JSON}
\newcommand{\jsonsch}{JSON Schema} 
\newcommand{\kw}[1]{\textbf{#1}}
\renewcommand{\kw}[1]{\ensuremath{\mathtt{#1}}}
\newcommand{\key}[1]{\ensuremath{\mathit{#1}}}
\newcommand{\akey}[1]{\ensuremath{\mathsf{#1}}}

\newcommand{\xnot}{\kw{not}}
\newcommand{\xtrue}{\kw{true}}
\newcommand{\xfalse}{\kw{false}}
\newcommand{\xnull}{\kw{null}}

\newcommand{\xany}{\kw{anyOf}}
\newcommand{\xall}{\kw{allOf}}
\newcommand{\xmin}{\kw{minimum}}
\newcommand{\xmax}{\kw{maximum}}
\newcommand{\xreq}{\kw{required}}

\newcommand{\xtype}{\kw{type}}
\newcommand{\xexmin}{\kw{exclusiveMinimum}}
\newcommand{\xexmax}{\kw{exclusiveMaximum}}
\newcommand{\xprops}{\kw{properties}}

\newcommand{\xpattProps}{\kw{patternProperties}}
\newcommand{\xminP}{\kw{minProperties}}

\newcommand{\xthen}{\kw{then}}
\newcommand{\xif}{\kw{if}}
\newcommand{\xelse}{\kw{else}}

\newcommand{\xaddProps}{\kw{additionalProperties}}

\newcommand{\xaddIts}{\kw{additionalItems}}

\newcommand{\xmof}{\kw{multipleOf}}

\newcommand{\xminL}{\kw{minLength}}
\newcommand{\xpatt}{\kw{pattern}}
\newcommand{\xuniqIt}{\kw{uniqueItems}}
\newcommand{\xcont}{\kw{contains}}

\newcommand{\xminC}{\kw{minContains}}
\newcommand{\xmaxC}{\kw{maxContains}}
\newcommand{\xminIt}{\kw{minItems}}
\newcommand{\xmaxIt}{\kw{maxItems}}
\newcommand{\xit}{\kw{items}}

\newcommand{\xdref}{\kw{\$ref}}

\newcommand{\xdefault}{\kw{default}}
\newcommand{\xdefs}{\kw{definitions}}
\newcommand{\xddefs}{\kw{\$defs}}

\newcommand{\xid}{\kw{id}}
\newcommand{\xdid}{\kw{\$id}}
\newcommand{\xda}{\kw{\$anchor}}

\newcommand{\ADefKey}{\akey{root}}
\newcommand{\DefKey}{\akey{def}}
\newcommand{\DefWithKey}[3]{{#1}\,{#2}={#3}}
\newcommand{\Def}[2]{\DefWithKey{\DefKey}{#1}{#2}}
\newcommand{\ADef}[2]{\DefWithKey{\ADefKey}{#1}{#2}}

\renewcommand{\Ref}[1]{\key{#1}}

\renewcommand{\comment}[1]{}
\newcommand{\hide}[1]{}
\newcommand{\code}[1]{}

\newcommand{\Iff}{\Leftrightarrow}
\newcommand{\Implies}{\Rightarrow}

\newcommand{\Or}{\vee}
\newcommand{\BigOr}{\bigvee}

\newcommand{\TypeOf}{\key{TypeOf}}

\renewcommand{\And}{\wedge}
\newcommand{\BigAnd}{\bigwedge}
\newcommand{\XOr}{\oplus}
\renewcommand{\XOr}{\circled{\text{\small 1}}}
\newcommand{\XOrVar}{\circled{\text{\small 1}}^*}
\newcommand{\NotXOr}{\circled{\text{\small \sout{1}}}^*}
\newcommand{\Not}{\neg}
\newcommand{\True}{{\bf t}}
\newcommand{\False}{{\bf f}}
\newcommand{\Type}{\akey{type}}
\newcommand{\Num}{\akey{Num}}
\newcommand{\Str}{\akey{Str}}
\newcommand{\Int}{\akey{Int}}
\newcommand{\Arr}{\akey{Arr}}
\newcommand{\Obj}{\akey{Obj}}
\newcommand{\Bool}{\akey{Bool}}
\newcommand{\Null}{\akey{Null}}
\newcommand{\TT}[1]{\Type({#1})}
\newcommand{\TNum}{\TT{\Num}}
\newcommand{\TStr}{\TT{\Str}}

\newcommand{\TInt}{\TT{\Int}}
\newcommand{\TArr}{\TT{\Arr}}
\newcommand{\TObj}{\TT{\Obj}}
\newcommand{\TBool}{\TT{\Bool}}
\newcommand{\TNull}{\TT{\Null}}
\newcommand{\Mof}{\akey{mulOf}}
\newcommand{\NotMof}{\akey{notMulOf}}
\newcommand{\Pat}{\akey{pattern}}
\newcommand{\NotPat}{\akey{notPattern}}
\newcommand{\Uni}{\akey{uniqueItems}}
\newcommand{\NotUni}{\akey{repeatedItems}}
\newcommand{\CProps}[1]{\akey{props}(#1)}
\newcommand{\Props}{\akey{props}}
\newcommand{\Ite}[2]{\akey{items}(#1;#2)}
\newcommand{\itdots}{\cdots}

\newcommand{\Bet}{\akey{betw}}
\newcommand{\XBet}{\akey{xBetw}}
\newcommand{\Len}{\akey{len}}
\newcommand{\Ex}{\#}
\newcommand{\Exi}{\exists}
\newcommand{\Pro}{\akey{pro}}
\newcommand{\Nam}{\akey{pNames}}
\newcommand{\ExNam}{\akey{exPName}}
\newcommand{\keykey}[1]{\key{\underline{#1}}}
\newcommand{\Con}{\akey{const}}
\newcommand{\Enu}{\akey{enum}}

\newcommand{\NotEnu}{\akey{notEnum}}
\newcommand{\IBT}{\akey{isBoolValue}}
\renewcommand{\IBT}{\akey{ifBoolThen}}
\newcommand{\Inf}{\infty}

\newcommand{\PReq}{\akey{pattReq}}
\newcommand{\APReq}{\akey{addPattReq}}
\newcommand{\OPReq}{\akey{orPattReq}}
\newcommand{\Req}{\akey{req}}
\newcommand{\AddP}{\akey{addp}}

\newcommand{\FalseP}{\False^{\cdot}}
\newcommand{\AndP}{\And^{\cdot}}

\newcommand{\NotP}{\Not^{\cdot}}
\newcommand{\SubP}{\subseteq^{\cdot}}

\newcommand{\FastSub}{\sqsubseteq}

\newcommand{\To}{\rightarrow}

\newcommand{\GG}[1]{\marginpar{\tiny g: {#1}}}

\newcommand{\M}{\ |\ }

\newcommand{\Sum}{\key{sum}}
\newcommand{\NotIf}{\key{NotIf}}

\newlength{\NL}
\newcommand{\Emptyset}{\emptyset}
\newcommand{\NotVarFun}{\key{co}}
\newcommand{\NotVar}[1]{\NotVarFun(\key{#1})}
\newcommand{\Sat}[2]{{#1}\vDash{#2}}

\newcommand{\NN}{\,\hat{}\,}
\newcommand{\Set}[1]{\{\,{#1}\,\}}
\newcommand{\SetTo}[1]{\{1..{#1}\}}
\newcommand{\SetFromTo}[2]{\{{#1}..{#2}\}}

\newcommand{\NR}[1]{\Ref{\NotVar{#1}}}

\newcommand{\nule}{\nullv}
\newcommand{\num}{n}
\newcommand{\nullv}{\ensuremath{\text{null}}}
\newcommand{\str}{s}
\newcommand{\V}{\CF{V}}
\newcommand{\CF}[1]{\mathit{#1}}
\newcommand{\J}{{J}}

\newcommand{\truev}{\ensuremath{\text{true}}}
\newcommand{\falsev}{\ensuremath{\text{false}}}

\newcommand{\semca}[2]{[\![ #1 ]\!]_{#2}}

\newcommand{\setst}[2]{\{ #1 \ | \ #2\}}
\newcommand{\rlan}[1]{L(#1)}
\newcommand{\noa}[1]{|#1|}

\newcommand{\TNBool}{\TG{\NBo}}
\newcommand{\TNArr}{\TG{\NAr}}
\newcommand{\TNNum}{\TG{\NNu}}
\newcommand{\TGIN}[1]{\akey{#1}}
\newcommand{\TG}[1]{\{\TGIN{#1}\}}
\newcommand{\Nl}{\TGIN{\SNl}}
\newcommand{\Bo}{\TGIN{\SBo}}
\newcommand{\Nu}{\TGIN{\SNu}}
\newcommand{\St}{\TGIN{\SSt}}
\newcommand{\Ob}{\TGIN{\SOb}}
\newcommand{\Ar}{\TGIN{\SAr}}

\newcommand{\NBo}{\TGIN{\overline{\SBo}}}
\newcommand{\NNu}{\TGIN{\overline{\SNu}}}

\newcommand{\NAr}{\TGIN{\overline{\SAr}}}
\newcommand{\TNl}{\TG{\SNl}}
\newcommand{\TBo}{\TG{\SBo}}
\newcommand{\TNu}{\TG{\SNu}}
\newcommand{\TSt}{\TG{\SSt}}
\newcommand{\TOb}{\TG{\SOb}}
\newcommand{\TAr}{\TG{\SAr}}
\newcommand{\SNl}{{Null}}
\newcommand{\SBo}{{Bool}}
\newcommand{\SNu}{{Num}}
\newcommand{\SSt}{{Str}}
\newcommand{\SOb}{{Obj}}
\newcommand{\SAr}{{Arr}}
\newcommand{\PP}[1]{\NN\,{#1}}
\newcommand{\PPP}[1]{\NN{#1}\$}
\renewcommand{\CProps}[1]{#1}
\newcommand{\Next}{\kw{next}}
\newcommand{\EOV}{\kw{EOV}}
\newcommand{\DOV}{\kw{DOV}}
\newcommand{\Pop}{\emph{Populated}}
\newcommand{\Open}{\emph{Open}}
\newcommand{\Emp}{\emph{Empty}}
\newcommand{\Sle}{\emph{Sleeping}}

\begin{document}

\title{Not Elimination  and Witness Generation for JSON Schema}


\author{Mohamed-Amine Baazizi}
\affiliation{%
  \institution{Sorbonne Universit\'e, LIP6 UMR 7606}
}
\email{baazizi@ia.lip6.fr}
\author{Dario Colazzo}
\affiliation{%
  \institution{Universit\'e Paris-Dauphine, PSL Research University}
}
\email{dario.colazzo@dauphine.fr}

\author{Giorgio Ghelli}
\affiliation{%
  \institution{Dipartimento di Informatica, Universit\`a di Pisa}
}
\email{ghelli@di.unipi.it}

\author{Carlo Sartiani}
\affiliation{%
  \institution{DIMIE, Universit\`a della Basilicata}
}
\email{carlo.sartiani@unibas.it}
\author{Stefanie Scherzinger}
\affiliation{%
  \institution{Universit{\"a}t Passau}
}
\email{stefanie.scherzinger@uni-passau.de}
%


\begin{abstract}
JSON Schema is an evolving standard for the description of families of JSON documents. JSON Schema is a logical language, based on a set of \emph{assertions} that describe features of the JSON
value under analysis and on logical or structural combinators for these assertions. As for any logical language, problems like  satisfaction,  not-elimination,  schema satisfiability, schema inclusion and equivalence, as well as witness generation, have both theoretical and practical interest.
While satisfaction is trivial, all other problems are quite difficult, due to the combined presence of negation, recursion, and
complex assertions in JSON Schema. To make things even more complex and interesting, JSON Schema is not algebraic, since we have both syntactic and semantic interactions between different keywords in the same schema object. 

With such motivations, we present in this paper an algebraic characterization of JSON Schema, obtained by adding opportune operators, and by mirroring existing ones. We present then  algebra-based  approaches for dealing with not-elimination and witness generation problems, which play a central role as they lead to solutions for the other mentioned complex problems.
\end{abstract}


\maketitle

\section{Introduction}

\subsection{Aim of the paper}

{\jsonsch} \cite{jsonschema} is an evolving standard for the description of families of {\json} documents.

{\jsonsch} is a logical language, based on a set of \emph{assertions} that describe features of the {\json}
value under analysis and on logical or structural combinators for these assertions. As for any logical language, the following
problems have a theoretical and practical interest:
\begin{compactitem}
\item satisfaction $\Sat{J}{S}$: does a {\json} document $J$ satisfy schema~$S$?
\item not-elimination: is it possible to rewrite a schema to an equivalent form without negation?
\item satisfiability of a schema: does a document $J$ exist such that  $\Sat{J}{S}$?
\item schema inclusion $S \subseteq  S^{\prime}$: does, for each document $J$,  $\Sat{J}{S} \Implies \Sat{J}{S^{\prime}}$?
\item schema equivalence $S \equiv  S^{\prime}$: does, for each document $J$, $\Sat{J}{S} \Iff \Sat{J}{ S^{\prime}}$?
\item witness generation: is there an algorithm to generate one element $J$ for any non-empty schema $S$?
\end{compactitem}

While satisfaction is trivial, all other problems are quite difficult, due to the combined presence of negation, recursion, and
complex assertions.

A second aspect that makes the task difficult is the non-algebraic nature of {\jsonsch}. 
A language is ``algebraic'' when the applicability and the semantics of its operators only depends on the semantics of their operands.
In this sense, {\jsonsch} is not algebraic, since we have both syntactic and semantic interactions between different
keywords in the same schema object, such as the prohibition to repeat a keyword inside a schema object, or the 
interactions between the
``\xprops'' and ``\xaddProps'' keywords.  
For instance, the following schema%
\footnote{Example taken from the \webref{https://github.com/json-schema-org/JSON-Schema-Test-Suite/blob/master/tests/draft6/additionalProperties.json}{JSON Schema Test Suite} (link available in the PDF).}
demands that any properties other than \verb!foo! and \verb!bar! must have boolean values.
\begin{Verbatim}[fontsize=\footnotesize,xleftmargin=5mm]
{ "properties": {"foo": {}, "bar": {}},
  "additionalProperties": {"type": "boolean"} }
\end{Verbatim}

Such features complicate the tasks of reasoning about the language and of writing code for its manipulation.

\subsection{Main contributions}
\subsubsection*{JSON Algebra}
We define a \emph{core algebra}, which features a subset of {\jsonsch}
assertions. This algebra is minimal,  that is, no operator can be defined starting from the others. 

\subsubsection*{Not elimination}

We show that negation cannot be eliminated from {\jsonsch}, since there are some assertions whose complement cannot be 
expressed without negation. 
We enrich the core algebra with primitive operators to express those missing 
complementary operators, and we give a \emph{not elimination} algorithm for the enriched algebra.
To our knowledge, this is the first paper where not elimination is completely defined, with particular regard to the treatment of negation and recursion.

\subsubsection*{Witness generation}
We define an approach  for witness generation for the complete {\jsonsch} language, with the only exception of the 
\xuniqIt\ operator, hence solving the satisfiability and inclusion problems for this sublanguage. 

For space reasons, many details and formal aspects presented in the complete report \cite{long} are not reported here, including the extension to \xuniqIt\  for witness generations. The presentation of several steps (especially for witness generation) is driven/based by/on examples.

Also, we would like to stress that results presented in this paper takes part of research activities \cite{long} that are still in progress. So our main aim here is to present existing results, mainly at the definition and formalisation level of algorithms. 

\subsection{Paper outline}

The rest of the paper is organized as follows. In Section \ref{sec:prelim} we briefly describe {\json} and {\jsonsch}, while in Section \ref{sec:algebra} we introduce our algebraic framework. In Section \ref{sec:notelimination}, then, we show how algebraic expressions can be rewritten so as to eliminate negation. In Section~\ref{sec:witness}, next, we discuss witness generation, while in Section \ref{sec:relworks} we analyze some related works. In Section \ref{sec:concl}, finally, we draw our conclusions.



\section{Preliminaries}\label{sec:prelim}

\subsection{JSON data model}

{\json} values are either basic values, objects, or arrays.
Basic values~$B$ include  the \nule\ value, booleans, numbers $\num$, and strings $\str$.
Objects represent sets of members, each member being a name-value pair~$(l,\V)$, and arrays represent ordered sequences of values.
We will use~$\J$ to range over {\json} expressions and $\V$ to range over the values denoted by such expressions, 
according to the semantics defined below, but
the two notions are so similar that we will often ignore this distinction.

We will only consider here objects without repeated names. 
In {\json} syntax, a name is itself a string, hence it is surrounded by quotes; for the sake of simplicity, we avoid these quotes in our notation, that is, we write   \{ name : ``John'' \} rather than  \{ ``name'' : ``John'' \}.


\[
\begin{array}{llrllllllll}
\J ::= & B \mid O \mid A                       & &\text{\bf {\json} expressions}  \\
B ::=	 & \nule \mid \truev \mid \falsev \mid \num \mid \str  &  & \text{\bf Basic values}\\
	&  \num\in\Num, \str\in\Str  \\ 
O ::= &  \{l_1:\J_1,\ldots,l_n:\J_n \}\    && \text{\bf Objects} \\
 &	\ n\geq 0, \ \  i\neq j \Rightarrow l_i\neq l_j \\
A ::= &  [\J_1, \ldots, \J_n ]\ \ \ & n\geq 0  &  \text{\bf Arrays} \\
\end{array}
\]

\begin{definition}[Value equality and sets of values] In the following we denote value equality with the usual notation $\J_1 = \J_2$, with the expected meaning on base values, while on Objects we have that  $O_1 = O_2$ if and only if $O_1= \{l_1:\J_1,\ldots,l_n:\J_n \}$ and  $O_2= \{l_{\pi(1)}:\J'_{\pi(1)},\ldots,l_{\pi(n)}:\J'_{\pi(n)} \}$ with $\pi$ a permutation over $I=\SetTo{n}$ and $\J_i=\J'_{\pi(i)}$ for each $i\in I $. On arrays we have  $A_1 = A_2$  if and only if $ A_1 = [\J_1, \ldots, \J_n ] $ and $A_2 = [\J'_1, \ldots, \J'_n ] $ with $J_i = J'_i$ for each $i\in \SetTo{n}$.

Sets of {\json} values are defined accordingly: a \emph{set} of {\json} values is a collection with no repetition with respect to this notion of equality, and two sets are equal when they have the same values with respect to this notion of equality.

\end{definition}

\subsection{JSON Schema}
{\jsonsch} is a language for defining the structure of {\json} documents. It is maintained by the Internet Engineering Task Force IETF \cite{ietf}. Its latest version has been produced on  2019-09 \cite{Version09} but is not widely used compared to the intermediate Draft-06. 

{\jsonsch} uses the {\json} syntax. Each construct is defined using a {\json} object with a set of fields describing assertions relevant for the values being described. Some assertions can be applied to any {\json} value type (e.g., \emph{type}), while others are more specific (e.g., \emph{multipleOf} that applies to numeric values only). 
The syntax and semantics of {\jsonsch} have been formalized in \cite{DBLP:conf/www/PezoaRSUV16} following the specification of Draft-04. 
We limit ourself to an informal discussion revealing the possible constraints associated to each kind of type:
\begin{itemize}
\item when defining a \emph{string}, it is possible to restrict its length by specifying the \emph{minLength} and \emph{maxLength} constraints and to define the \emph{pattern} that the string should match;  
\item when defining a \emph{number}, it is possible to define its range of values by specifying  any combination of \emph{minimum} / \emph{exclusiveMinimum} and \emph{maximum} / \emph{exclusiveMaximum}, and to define whether it should be \emph{multipleOf} a given number;
\item when defining an \emph{object}, it is possible to define its \emph{properties},  the type of its \emph{additionalProperties} and the type of the properties matching a given pattern (i.e. \emph{patternProperties}). It is also possible to restrict the minimum and maximum number of properties using \emph{minProperties} and \emph{maxProperties}, and to indicate which properties are \emph{required};
\item when defining an \emph{array}, it is possible to define the type of its \emph{items} and the type of the \emph{additionalItems} which were not already defined by \emph{items}, and to restrict the minimum and maximum size of the array; moreover, it is also possible to enforce unicity of the items using \emph{uniqueItems}.
\end{itemize}
{\jsonsch} allows for combining assertions using standard boolean connectives: \emph{not} for negation, \emph{allOf} for conjunction, \emph{anyOf} for disjunction, and \emph{oneOf} for exclusive disjunction.
Moreover, indicating the set of accepted values can be done using the \emph{enum} constraint.

\section{The algebra}\label{sec:algebra}


We opt for a core algebra that is based on a minimal set of operators expressive enough to capture all {\jsonsch} constraints, including those of the last 2019 specification \cite{Version09}. 
We consider two variants of this algebra: one variant making explicit use of negation and another variant where negation is substituted with a set of operators expressing negation implicitly.
The syntax of the two algebras is presented in Figure \ref{fig:core-syntax}. 

\begin{figure}[ht]
$$
\begin{array}{llll}
T & ::= & \Arr \M \Obj \M \Null \M \Bool \M \Str  \M \Num \\[\NL]
r & ::= &  \text{\em JSON Schema regular expression} \\[\NL]
b & ::= & \xtrue \M \xfalse \\[\NL]
S & ::= & \IBT(b)\M \Pat(r) \M \Bet_{m}^{M}  \M \Mof(n)   \\[\NL]
&&  \M \Pro_{i}^{j} \M \key{r} : S    
 \M i \D j : S \M \Ex_{i}^{j}S \M  \Uni  \\[\NL]
&& 
 \M  \ADef{x_1}{S_1} , \Def{x_2}{S_2}, \ldots, \Def{x_n}{S_n} \M \Ref{x} \M \ S_1 \And S_2 \\[2\NL]
\multicolumn{2}{r}{\text{either:}}&  \M \Not S \\[2\NL]
\multicolumn{2}{r}{\text{or:}}&
\M \NotPat(r)  \M \XBet_{m}^{M}  \M \NotMof(n) \\[\NL]
&&
\M \PReq(r : S) \M \NotUni \\[\NL]
&&
\M S_1 \Or S_2 \M \Type(T)  \\[\NL]
\end{array}
$$
\caption{Syntax of the core algebras.}
\label{fig:core-syntax}
\end{figure}

In $\Mof(n)$, $n$ is a number.
In $\Bet_{m}^{M}$ and in $\XBet_{m}^{M}$
$m$ is either a number or $-\Inf$, $M$ is either a number or $\Inf$. 
In $ \Pro_{i}^{j}$ and in $\Ex_{i}^{j}S$, $i$ is an integer with $i \geq 0$,
while  in $i \D j : S$,  $i$  is an integer with $i \geq 1$.
In these three operators, $j$ is either an integer with the same lower bound as $i$,
or $\Inf$. 

This algebra features two possibilities for the negation: the \emph{core algebra with $\Not$}, which explicitly uses negation $\Not S$, and the \emph{not-eliminated core algebra}, in which $\Not S$ is substituted by the 
seven operators of the last three lines.

We show below that negation can express the seven operators of the \emph{not-eliminated core algebra}, and then  we prove the opposite direction, that is, the fact that negation can be
eliminated using these operators.

\begin{remark}
In this paper we will assume that {\jsonsch} regular expressions are indeed regular expressions,
hence they are closed under negation and intersection, and these operations are decidable.
This is actually good enough in practice, but is not true in general \cite{DBLP:journals/mst/Freydenberger13}.
\end{remark}

All operators that are related to one specific type, that is, all operators
in the first and second line, have an implicative semantics, where the condition is
always: ``if the instance belongs to the type associated with this assertion''.
We say that they are \emph{implicative typed assertions} (ITEs).

The meaning of each operator is informally given as follows: 
\begin{itemize}
\item $\IBT(b)$  means: \emph{if} the instance is a boolean, \emph{then} it is~$b$.
\item $\Pat(r)$ means: \emph{if} the instance is a string, \emph{then} it matches~$r$.
\item $\Bet_{m}^{M}$ means: if the
instance is a number, \emph{then} it is included between $m$ and $M$, extreme included.
%
\item $\Mof(n)$ means: if the instance is a number, \emph{then} it is a multiple of $n$.
%
\item $\Pro_{i}^{j}$ means: if the instance is an object, \emph{then} it has at least $i$ properties and at most $j$.
\item The assertion $\key{r} : S$ is two times implicative, since it means: if the instance is an object and if $k$ is a name of this object that
matches the pattern $r$, \emph{then} the value associated with $k$ satisfies $S$.  Hence, it is satisfied by any instance that is not an object and also by any object where
no name matches $r$. 
\item $\PReq(r : S)$ means: if the instance is an object, \emph{then} it contains at least one name that matches $r$ and whose value matches $S$.
\item The assertion $i \D j : S$ is two times implicative, and it means: if the instance is an array and if 
$J$ is an element in a position $l$ such that $i \leq l \leq j$, \emph{then} $J$ satisfies $S$.  Hence, it is satisfied by any instance that is not an array and also by any array that is shorter than $i$, such as the empty array.
It does not constrain in any way the elements of position less than $i$ or greater than $j$,
nor does it force any position between $i$ and $j$ to be actually used.
In {\jsonsch} this assertion is expressed as 
$\xit : [ \xtrue^1,\ldots,\xtrue^{i-1},S^i,\ldots,S^j]$ when $j$ is finite,
and $\{ \xit : [ \xtrue^1,\ldots,\xtrue^{i-1}], \xaddIts : S\}$ when $j=\Inf$.
\item The assertion $\Ex_{i}^{j}S$ means that, if the instance is an array, \emph{then} the total number of elements that satisfy $S$ is included between
$i$ and~$j$. The operator 
$\Ex_{i}^{j}S$ corresponds to the combination of $\xcont$ with $\xminC$ and $\xmaxC$ that has been introduced by version 2019-09. Prior to this version, the 1-to-$\Inf$ form
$\Ex_{1}^{\Inf}S$ could be expressed in {\jsonsch} using $\xcont$, 
while the restricted version
$\Ex_{i}^{j}\True$ could be expressed using $\xminIt$ and $\xmaxIt$,
but the general form was not available.
For example, one can prove that the assertions $\Ex_{2}^{\Inf}\Type(\Int)$ and 
$\Ex_{0}^{2}\Type(\Int)$ cannot be expressed in {\jsonsch} Draft-06. 
We omit the upper-bound when it is $\Inf$ and write $\Ex_{i}$ instead of $\Ex_{i}^\Inf$. 
\item The assertion $\Uni$ means that, if the instance is an array, \emph{then} all of its items are distinct.
\item $\Not S$ is satisfied iff $S$ is not satisfied.
\item $S_1 \And S_2$ is satisfied when both $S_1$ and $S_2$ are satisfied.
\item 
Finally, $\ADef{x_1}{S_1} , \Def{x_2}{S_2}, \ldots, \Def{x_n}{S_n}$ defines $n$ mutually recursive variables, so that $\Ref{x}_i$ can
be used as an alias for $S_i$ inside any of $S_1 , \ldots, S_n$. One and only one variable is defined
as \emph{active} using $\ADef{x_1}{S_1}$, and an instance satisfies this assertion iff it satisfies the active 
variable.
\end{itemize}

We require recursion to be \emph{guarded}, according to the following definition: let us say that $x_i$ 
\emph{directly depends} on $x_j$ if $x_j$
appears in the definition of $x_i$ under a chain of boolean operators. 
For example, in $\Def{x}{(\key{r}:\Ref{y}) \And  \Ref{z} }$,
$x$ directly depends on $z$, but not on~$y$.
Recursion is not guarded if the transitive closure of this relation contains
a reflexive pair $(x,x)$. Informally, any cyclic dependency must traverse a \emph{typed} operator, that is, one that is different from $\Not$,
$\And$ and $\Or$.

\medskip

The semantics of a schema $S$ is the set of JSON instances  $\semca{S}{e}$
that \emph{satisfy} that schema, as specified below; the $e$ parameter is used to interpret variables, and will
be explained later.

In the semantics below,  $\rlan{r}$ denotes the regular language generated by $r$,  while $\noa{R}$ is  the number of top-level members of the object $R$. 

Universal quantification on an empty set is true, and the set $\SetTo{0}$ is empty, so that, for example,
both ${  i \D j : S }$ and $\Uni$ hold on the empty array.

\begin{figure}[ht]
\[
\begin{array}{lcl}
   \semca{\IBT(b)}{e} & = & \setst{J}{J \textit{ is a boolean } \Rightarrow  J=b }\\
   \semca{\Pat(r)}{e} & = & \setst{J}{J \textit{ is a string } \Rightarrow J \in  \rlan{r}} \\
   \semca{\Bet_{m}^{M}}{e} & = & \setst{J}{ J \textit{ is a number } \Rightarrow  m \leq J \leq M } \\
  \semca{\Mof(n)}{e} & = & \setst{ J}{ J \textit{ is a number } \Rightarrow  \\
     &&
\exists k \textit{ integer  with } J = k*n} \\
    \semca{\Pro_{i}^{j} }{e} & = & \setst{J}{J \textit{ is an object}  \Rightarrow  i \leq \noa{J} \leq j} \\
     \semca{\key{r} : S  }{e} & = & \setst{J}{J \textit{ is an object}   \Rightarrow\\
     &&
 ((l:\J') \in J \wedge  l\in\rlan{r}) \ \Rightarrow J' \in \semca{S}{e} } \\
\semca{  i \D j : S }{e} & = & \setst{J}{ J =  [\J_1, \ldots, \J_n ] \Rightarrow  \\
     &&
          \forall p\in\SetTo{n} \cap \SetFromTo{i}{j}  \Rightarrow J_p \in   \semca{  S }{e} }\\
\semca{  \Ex_{i}^{j}S  }{e} & = & \setst{J}{ J =  [\J_1, \ldots, \J_n ] 
          \Rightarrow \\
     &&
i \leq \ |\setst{l}{\J_l\in \semca{S}{e}}|\ \leq j }\\
\semca{ \Uni \  }{e} & = & \setst{J}{ J =  [\J_1, \ldots, \J_n ] \Rightarrow \\
     &&
 \forall i,j \in \SetTo{n}. \ i\neq j \ \Implies \ J_i \neq J_j}\\
 \semca{S_1 \And S_2}{e} & = &  \semca{S_1 }{e}  \cap   \semca{ S_2}{e} \\
\semca{\Not S}{e}  & = & \setst{J}{ J \not\in \semca{S}{e}  } \\
 \semca{\Ref{x}}{e}  &= & \semca{e(x)}{e}\\
\multicolumn{3}{l}{
\semca{\ADef{x_1}{S_1} , \ldots, \Def{x_n}{S_n}}{e} \ =\ 
\semca{S_1}{e,x_1\To S_1,\ldots,x_n\To S_n}
}  \\
 \end{array}
\]

\caption{Semantics of the algebra with explicit negation.}
\label{tab:corealg-sem}
\end{figure}

{\jsonsch} specification says that the ``result'' of a variable $\Ref{x}$ is the ``result'' of the referenced schema, 
which may be formalized as follows: an equation system is equivalent to the first element, evaluated in an environment
$e$ where every variable is associated to its definition. When a variable is met, it is substituted by its definition in the 
current environment. We assume that in $e,x\To S$ the new binding $x\To S$ hides any previous binding for $x$, so that
we support the usual rule that variables in an inner scope hide variables in an outer scope.


This formalization mirrors {\jsonsch} specifications, but it is not totally satisfactory since, in clause for $ \semca{\Ref{x}}{e}$, the right hand side 
is actually bigger than the left hand side $ \semca{e(x)}{e}$, hence this is not an inductive definition, and actually it may be the case that we have different interpretations of the $\semca{}{}$ function that are compatible with that equation, or that we have none.
This is a classical problem that we solve in the classical way (see \cite{long}), by resorting to a least-fixed-point semantics. 

Hereafter will we often use the derived operators $\True$ and $\False$. 
$\True$ stands for
``always satisfied'' and can be expressed, for example, as $\Pro_0^{\Inf}$, which is satisfied by any instance.
$\False$ stands for ``never satisfied'' and can be expressed, for example, as $\Not \True$.

\subsection{Semantics of the negated operators}

As we said before, the seven operators in the last group are redundant in presence of negation.
$\Type(T)$ corresponds to the {\jsonsch} operator $\xtype$, 
$\XBet$ corresponds to $\xexmin$ and $\xexmax$, and $\Or$ corresponds to $\xany$.
The other three
do not correspond to {\jsonsch} operators, but can still be expressed in {\jsonsch}, through the
negation of $\xmof$, $\xpattProps$, and $\Uni$.

We now show how these operators can be expressed in the core algebra with $\Not$.

The operator $\Type(T)$ can be expressed using negation as follows.
In all cases, we use the implicative nature of typed operators.
For example, in $\Pat(\ \hat{}\ \$) \And \Pat(.) $ we have chosen two patterns with
empty intersection. No string may satisfy this conjunction, but any instance that is not a string
would satisfy it, hence it identifies instances that are not strings. Hence, its complement identifies
instances that are strings. All other types are defined in the same way, using two typed assertions that
are not compatible, hence are only satisfied by instances of any other type. 
The only exception is the $\Null$ type, since we have no typed operators for $\Null$, hence we take the 
complement of the other five types.
$$\begin{array}{lll}
\Type(\Str) &=& \Not ( \Pat(\ \hat{}\ \$) \And (\Pat(.) ) \\[\NL]
\Type(\Num) &=& \Not ( \Bet_0^0 \And \Bet_1^1  ) \\[\NL]
\Type(\Bool) &=&\Not (\IBT(\xtrue) \And \IBT(\xfalse)) \\[\NL]
\Type(\Obj) &=& \Not ( \Pro_{0}^{0} \And\Pro_{1}^{1} ) \\[\NL]
\Type(\Arr) &=& \Not ( 1\D 1 : \False \And \Ex_1^{\Inf}\True   ) \\[\NL]
\Type(\Null) &=& \Not \Type(\Str) \And \Not \Type(\Num) \And \Not \Type(\Bool)  \\[\NL]
                         &&\qquad \And \Not \Type(\Obj) \And \Not \Type(\Arr)  \\[\NL]
\end{array}$$
The other six operators can be expressed as follows, where we use the $\Type$ operator
and $\Implies$ for readability.
In order to stay in the core algebra, $\Type(T) \Implies S$ should be
written as $\Not \Type(T) \Or S$, and $\Not \Type(T)$ should be expressed as in the
table above. Observe that the semantics of the negative operators is implicative, exactly as
that of the positive operators: they are always satisfied by any instance that does not belong to the
associated type.
$$\begin{array}{lll}
\NotPat(r)  & = & \TStr \Implies \Not \Pat(r) \\[\NL]
\XBet_m^M &=&\Type(\Num) \Implies  (\Not \Bet_{-\Inf}^m \And \Not \Bet_M^{\Inf})\\[\NL]
\NotMof(n) &=&\Type(\Num) \Implies \Not \Mof(n)  \\[\NL]
\PReq(r : S) &=&\Type(\Obj) \Implies \Not (\key{r} : \Not S)  \\[\NL]
\NotUni &=& \Type(\Arr) \Implies \Not \Uni  \\[\NL]
S_1 \Or S_2 &=& \Not (\Not S_1 \And \Not S_2) \\[\NL]
\end{array}$$

The definition of $\PReq(r : S)$ deserves an explanation. 
The implication $\Type(\Obj) \Implies \ldots$ just describes its implicative nature --- it is satisfied by any
instance that is not an object. Since $\key{r}:\Not S$ means that, if a name matching $r$ is present, then its value satisfies~$\Not S$, any instance that does not satisfy $\key{r}:\Not S$ must possess a member name that matches $r$ and 
whose value does not satisfy~$\Not S$, that is, satisfies $S$. Hence, we exploit here the fact that the negation of an implication
forces the hypothesis to hold.

\comment{
For reference, we also give a semantics for the derived operators
\begin{table}[ht]
\[		
\begin{array}{lcl}
 \semca{\Type(\Null)}{e} & = &\{\nule\} \\
  \semca{\Type(\Bool)}{e} & = & \{\truev, \ \falsev \}\\
   \semca{\Type(\Arr)}{e} & = & \setst{J}{J \textit{ is an array}} \\
    \semca{\Type(\Obj)}{e} & = & \setst{J}{J \textit{ is an object}} \\
    \semca{\Type(\Num)}{e} & = & \setst{J}{J \textit{ is a number  }} \\
\semca{\Type(\Str)}{e} & = & \setst{J}{J \textit{ is a string }} \\
     \semca{\NotMof(n)}{e} & = & \setst{ J}{ J \textit{ is a number }       \Rightarrow
     \\
     &&
    \not \exists k \textit{ integer  with } J = k*n } \\
\semca{ \NotUni \  }{e} & = & \setst{J}{ J =  [\J_1, \ldots, \J_n ] \Rightarrow \\
     &&
 \exists i,j \in \SetTo{n}. \ i\neq j \ \wedge \ J[i]=J[j]}\\
\semca{ \PReq(r : S)}{e} 
 & = & \setst{J}{J \textit{ is an object}  \Rightarrow  \\
     &&
 \exists (l:\J') \in J. \  l\in\rlan{r} \wedge \J' \in \semca{S}{e}}\\
\semca{S_1 \Or S_2}{e} & = &  \semca{S_1 }{e}  \cup   \semca{ S_2}{e} 
 \end{array}
\]
\caption{\label{tab:sem} Semantics of the algebra with implicit negation.}

\end{table}
 }

\begin{remark}
The negation of the operator $i_1 \D i_2 : S$ can be expressed with no need of
a specific negative operator, since it can be expressed using the same operator plus
$\Ex_n^{\Inf}\True$.
The negation of $i \D \Inf : S$ can also be expressed with no need of
a specific negative operator, since it can be expressed using an exponential number of
$i_1 \D i_2 : S$ and $\Ex_n^{\Inf}\True$, plus the operator $\Ex_n^{\Inf}S$,
as described in Section \ref{sec:notitem}.

Without the $\Ex_n^{\Inf}S$ operator, which is our representation of the combination of \xcont,
\xminC\ and \xmaxC\ (operators which were introduced in {\jsonsch} 2019), 
in order to express the negation of $i \D \Inf : S$
we would need at least an operator $\Exi( i \D j : S)$ that specifies that at least one element 
between $i$ and $j$ matches $S$. This operator would be strictly less expressive than 
the $\Ex_n^{\Inf}S$ operator, but this is explained in the next remark.
\end{remark}

\begin{remark}
While $\Ex_1^{\Inf} S$ can be immediately translated as 
$\TArr \Implies \Not (1 \D \Inf : \Not S)$, the assertions $\Ex_{min}^{\Inf} S, min > 1$
and $\Ex_{0}^{Max} S, 0 < Max < \Inf$ cannot be expressed for any non trivial $S$ without the $\Ex$ operator.
\comment{
\begin{proof}
Assume, by absurd, that $S'$ is equivalent to $\Ex_0^{n} S$ with $0<n<\Inf$ and
with $S$ non-trivial. Let $k$ be the biggest $j$ less
than $\Inf$ such that $ i \D j : S''$ or $ j+1 \D \Inf : S''$ appears in $S'$.
The index $k$ is crucial: by the semantics of the core algebra, no assertion in $S'$ can
distinguish between two arrays where the \emph{set} of items that are found from
position $k+1$ onward is the same. 
Let $J$ be an element of $S$ and $J'$ and element of $\Not S$.
Consider the array that contains $k$ zeros followed by $n$ copies of $J$ and two 
copies of $J'$:
$A=[0^1,\ldots,0^{k},J^1,\ldots,J^n,J',J']$ and the array $A'$ 
where the $J'$ in position $n+1$ is substituted by $J$: 
$A'=[0^1,\ldots,0^{k},J^1,\ldots,J^n,J,J']$.
Since $S'$  is equivalente to $\Ex_0^{n} S$, we have that $A \in \semca{S}{e}$
and $A' \not\in \semca{S}{e}$.
This is impossible, since the two arrays have the same set $\{ J, J' \}$ of items in the positions
from $k+1$ onwards, hence no assertion inside $S'$ can accept one and refuse the other.
\end{proof}
}
\end{remark}


\subsection{Representing definitions and references}

{\jsonsch} defines a $\xdref : \key{path}$ operator that allows any subschema of the current schema to be referenced, as well as any subschema of a 
different schema that is reachable through a URI, hence implementing a powerful form of mutual recursion. 
The path $\key{path}$ may navigate through the nodes
of a schema document by traversing its structure, or may retrieve a subdocument on the basis of a special $\xid$, $\xdid$, or 
$\xda$ member ($\xda$ has been added in Version 2019/09), which can be used to associate a name to the 
surrounding schema object.
Despite this richness of choices, in most situations, according to our collection of JSON schemas, 
the subschemas that are referred are either the entire schema or those that are
collected inside the value of a top-level \xdefs\ member.
Hence, we defined a referencing mechanism that is powerful enough to translate every collection of 
JSON schemas, but that privileges a direct translation of the most commonly used mechanisms.
\hide{\footnote{This structural pattern also holds for most of the examples in the {\jsonsch} Test
Suite (e.g., \href{https://github.com/json-schema-org/JSON-Schema-Test-Suite/blob/master/
tests/draft6/ref.json}), an online collection of small JSON Schemas and sample documents
illustrating valid and invalid JSON documents.}}


In our referencing mechanism, a schema may have the following structure, where the schema $S_1$ associated
to the $\ADefKey$ definition is the one that is used to validate the instance, and every variable $x_i$ is bound to 
$S_i$.
$$
\ADef{x_1}{S_1} , \Def{x_2}{S_2}, \ldots, \Def{x_n}{S_n}
$$
Such a schema corresponds to a JSON schema whose root contains $S_1$ plus a \xdefs\ member, which in turn contains
the definitions $\Def{x_2}{S_2}, \ldots, \Def{x_n}{S_n}$. In {\jsonsch} the entire schema can be denoted as 
$\#$, but we preferred an explicit naming mechanism $\ADef{x_1}{S_1}$ for uniformity.

For example, this {\jsonsch} document:
$$
\{ a_1 : S_1, \ldots, a_n : S_n, 
    \xdefs : \{ x_1 : S'_1, \ldots, x_m : S'_m \}
\}\quad  (1)
$$
corresponds to the following expression, where $\underline{S}$ is the translation of $S$:
$$
\ADef{r}{\underline{\{ a_1 : S_1, \ldots, a_n : S_n\}}},\,
\Def{x_1} {\underline{S'_1}} , \ldots, \Def{x_m}{\underline{S'_m}} \quad  (2)
$$

This mechanism is as expressive as the combination of all {\jsonsch} mechanisms, at the price of some code duplication.
In order to translate any JSON document that uses references, in our implementation we first collect all 
paths used in any $\xdref : \key{path}$ assertion. Whenever \emph{path} is neither $\#$ nor 
\xdefs/\kw{k} for some $k$, we retrieve the referred subschema and copy it inside the \xdefs\ member
where we give it a name \emph{name}, and we substitute all occurrences of $\xdref : \key{path}$ with 
$\xdref : \xdefs/\kw{name}$, until we reach the shape (1) above.
While in principle this may increase the size of the schema from $n$ to 
$n^2$, in case we have paths that refer inside the object that is referenced by another path, in practice
we observed a factor of 2 in the worst cases.
When we have a collection of documents that refer one inside the other, we first merge the documents together
and then apply the same mechanism. It would not be difficult to extend the naming mechanism in order to 
avoid merging the documents, but we consider this extension out of the scope of this paper.

In our syntax, the $\ADef{x_1}{S_1} , \Def{x_2}{S_2}, \ldots, \Def{x_n}{S_n}$ construct is a first class
assertion, hence can be nested.
We defined the syntax this way just for uniformity, but in practice we only use schemas where 
this construct is only used at the outermost level.

\comment{
\begin{verbatim}
{ type : object ,
  properties : {...}
  definitions : {
      a1: S1,
      an: Sn
      }
}
\end{verbatim}

Becomes:

\begin{verbatim}
def root = { type : object , properties : {...} }
def a1: S1,
def an: Sn
\end{verbatim}

}


\subsection{From \jsonsch\/ to the algebra}
Our algebra expresses all assertions of Draft 6.  
The translation rules from \jsonsch\/ to our algebra are provided in Table \ref{tab:transl} by omitting symmetric cases (e.g.  ``\xmax'' : M, ``\xexmax''~: M, etc) which can be easily guessed.

\begin{table}[ht]
\begin{tabular}{l|l}
 ``\xmin'' : m 
& $\Bet_m^{\Inf} $\\
``\xexmin'' : m 
& $\XBet_m^{\Inf} $\\
``\xmof'' : n
& $\Mof(n)$\\[\NL]
``\xminL'' : m 
&  $\Pat(\,\hat{}\ .\{m,\}\,\$)$ \\ 
``\xpatt'': r 
 &  $\Pat(r)$\\[\NL]
``\xuniqIt'' : n
&  $\Uni$\\
``\xminIt'' : m
&  $\Ex_m^{\Inf} \True $\\
``\xcont'' : $S$, ``\xminC'' : m &$\Ex_m^{\Inf} \Tr{S}$ \\
``\xit'' : [$S_1$, \ldots, $S_n$], \\
\qquad \qquad ``\xaddIts'' : $S'$
&  $\Ite{\Tr{S_1} \itdots  \Tr{S_n}}{ \Tr{S'}}$\\
``\xit'' : $S$
&  $\Ite{}{\Tr{S}} $\\[\NL]
``\xminP'' : m 
&  $\Pro_m^{\Inf}$\\
``\xreq'' : [ \key{k_1},\ldots,\key{k_n} ]
&$ \TObj \Implies
( \Not ( \keykey{k_1} : \False) \And \ldots $ \\[\NL]
& \qquad \qquad \qquad $\And \Not (  \keykey{k_n} : \False ) ) $\\[\NL]

``\xprops'' : \{ i=1..n  \key{k_i} : $S_i$ \},
& $ \keykey{k_1} : \Tr{S_1},..,\keykey{k_n} : \Tr{S_n}$,    \\
\quad``\xpattProps'' : \\ 
\qquad \{i=1..m \key{r_i} :\,$PS_i$ \},
& $  \key{r_1} : \Tr{PS_1},..,\key{r_m} : \Tr{PS_m}$,  
\\
\quad``\xaddProps'' : $S$
&  $\NotP(\keykey{k_1}|\ldots|\keykey{k_n}|r_1|\ldots|r_m) : S$ \\[\NL]
\\[\NL]
\key{k} : J , 
 & $ \ADef{xroot} {\Tr{\key{k} : J}}$ \\
``\xdefs'' : \{ \key{x_1} : $S_1$, \ldots, \key{x_n} : $S_n$\},
 & $\Def{x_1}{\Tr{S_1}} , $\\
&  \qquad\qquad $\ldots, \Def{x_n}{ \Tr{S_n}}$\\
``\xdref'' : ``\#/\xdefs/$x$'' &   $\Ref{x}$ \\
``\xdref'' : ``\#'' &   $\Ref{\key{xroot}}$ \\
\end{tabular}
\caption{\label{tab:transl}Translation rules for \jsonsch}
\end{table}


The translation of $\Con(J)$, which is not described in Table \ref{tab:transl}, depends on the kind of $J$. For example, 
 when $J$ is a number $n$,  $\Con(J)$ is translated to $ \Type(\Num) \And \Bet_n^n  \  n\in \TNum$. The remaining cases are presented in the long version \cite{long}.
The rules for "\xall", "\xany" and "\xnot" are omitted and translation  of ``\xif'' : $S_{1}$, ``\xthen'' : $S_{2}$, ``\xelse'' : $S_{3}$  is  $(\Tr{S1}\And \Tr{S2})\Or(\Not{\Tr{S1}}\And \Tr{S3})$.

The rule for ``\xprops'' is rather intuitive: each $\key{k_i} : S_i$ is mapped to $\keykey{k_i} : \Tr{S_i}$ where $\keykey{k}$ is the regular expression $\key{\hat{}\ k\,\$} $ and $\NotP$ the negation operator of regular expressions. 

Concerning array descriptions, in Table \ref{tab:transl} and hereafter we adopt the  n-ary operator $\Ite{S_1 \itdots  S_n}{S_{n+1}}$, encoded over our algebra as follows. 
$$
\begin{array}{lllll}
\Ite{S_1 \itdots  S_n}{S_{n+1}} 
         &=& 1\D 1 : S_1, \ldots, n\D n : S_n,
                                             n+1\D \Inf : S_{n+1} \\[\NL]
\end{array}
$$

Observe that it preserves
the doubly implicative nature of the base operator, hence it is satisfied by any value that it is not an array, and it
does not force the array to actually contain any element, but it only specifies the schemas of elements if they are present.




\section{Negation elimination}\label{sec:notelimination}

We use not-elimination to indicate the property of a logic to express the negation of every formula
with no use of the negation operator. 

 
{\jsonsch} does not enjoy not-elimination, since it contains some assertions whose negation
cannot be expressed without a negation operator, such as $\xmof$ or $\xuniqIt$. 
Our algebra with the implicit negated operators is sufficient to rewrite any explicitly negated operator by eliminating the $\Not$.

%


\comment
{
\begin{theorem}
The positive core algebra cannot express $\NotUni$.
\end{theorem}

\begin{proof}
Assume, for a contradiction, that we can express the following type.
$$
\Ite(;\Type(\Int)) \And \NotUni
$$
Hence, there exists a schema $S$ written in the positive core algebra such that, for any number $n$, the sequence
$1,\ldots,n,n$ is allowed while $1,\ldots,n$ is not. Let $J$ be the biggest finite $j$ such that either $i:j : S'$ or $j:\Inf;S$
is present in $S$ and consider the sequences $1,\ldots,J+1$ and $1,\ldots,J+1,J+1$\ldots
To be completed keeping into account also $\Ex_i^j S$. 
\end{proof}
}

We present an algorithm to push negation down the syntax tree of any schema.
The algorithm proceeds in three phases:
\begin{enumerate}
\item Not-completion of variables: for every variable $\Def{x_n}{S_n}$ we define a 
corresponding $\Def{not\_x_n}{\Not S_n}$
\item Not-rewriting: we rewrite every expression $\Not S$ into an expression
where the negation has been pushed inside.
\end{enumerate}

We first present not-completion and not-rewriting for the $\Ite{}{}$ case,
and we then present the rest of not-rewriting.
  
\subsection{Not-completion of variables}\label{sec:notcompletion}

Not-completion of variables is the operation that adds a variable
$\key{not\_x}$ for every variable $\key{x}$ as follows:
$$\begin{array}{lllll}
\text{not-completion}(\ADef{x_0}{S_0} , \ldots, \Def{x_n}{S_n}) = \\[0.8ex]
\quad\ADef{x_0}{S_0},
, \ldots, \Def{x_n}{S_n}, \\[0.8ex]
\quad\Def{not\_x_0}{\Not S_0} , \ldots, 
\Def{not\_x_n}{\Not S_n}
\end{array}$$

After not-completion, every variable has a complement variable defined in the obvious way:
$\NotVar{x_i}=not\_x_i$ and $\NotVar{not\_ x_i}=x_i$. 
The complement $\NotVar{x}$ will later be used for not-elimination.

%
%
\subsection{Inversion of \Ite{}{}}\label{sec:notitem}

The inversion of 
$\Ite{S_1 \itdots S_n}{S}$ is the most complex case of not elimination.
According to its semantics, only a non-empty array may not satisfy that assertion. 
For instance,  the empty array as well as the string ``foo'' satisfy both $\Ite{\False}{\True}$
and $\Ite{}{\False}$, while any array with length 1 or more would violate both types.

More generally, we have the following formula
$$\Not\Ite{S_1 \itdots S_n}{S} =\TArr \And (N_1 \Or \ldots \Or N_n \Or N_{n+1})$$

that expresses the fact that an array $[J_1,\ldots,J_m]$ 
may not satisfy $\Ite{S_1 \itdots S_n}{S}$  in one of the following $n+1$ ways:
\begin{itemize}
\item The array has at least $1\leq i \leq n$ elements and element $J_i$ does not satisfy $S_i$:
$$N_i = \Ex_i^{\Inf}\True \And \Ite{\True_1,\ldots,\True_{i-1},\Not S_i}{\True}$$


\item The array has at least n+1 elements and some element $J_{n+l}$, with $l>0$,
does not satisfy the tail schema  $S$. This case $N_{n+1}$ is the most complex one and deserves some preliminary discussion.

\end{itemize}

%


Concerning $N_{n+1}$, we consider first  the following special cases.

\begin{enumerate}
\item 
The most common case is when $n=0$. In this case one single non-$S$ element is enough to 
violate $\Ite{}{S}$, hence we have
$N_{n+1}=\Ex_1^{\Inf}\Not S$, and the initial sequence 
$N_1 \And \ldots \And N_n$
is empty.\footnote{This case arises from the translation of \xit : $S$ where $S$ is not an array and
also in those rare situations where \xaddIts\ : $S$ is present and \xit\ is absent}

\item 
The second most common case is the one with $n>0$ and  $S=\False$. 
In this case, we violate the tail condition whenever the array has at least
$n+1$ elements, 
hence we have that $N_{n+1}=\Ex_{n+1}\True$.
\item The third most common case is $n>0$ and  $S=\True$. In this case, the tail condition cannot be violated, hence 
$N_{n+1}=\False$.
\comment{, or, equivalently, we just remove this case from the disjunction above for $\Not\Ite{S_1 \itdots S_n}{S}$. We generalise this case to the situation where $S$ includes every $S_i$. Again, in any array that violates 
$\Ite{S_1 \itdots S_n}{S}$ but does not satisfy $N_i$ for any $i$, that is, its first $n$ elements satisfy the
corresponding $S_i$, we have that the first $n$ elements all satisfy $S$, as a consequence of the
inclusion hypothesis, hence none of them satisfies $\Not S$.
Hence, the array violates $\Ite{S_1 \itdots S_n}{S}$ with an elements that has a position greater
than ${n+1}$ iff at least one element satisfies $\Not S$,
hence we have $N_{n+1}=\Ex_1\Not S$, that reduces to $\False$ when $S=\True$.}
%
\item 
A last special case is that where the array schema has length 1, that is $\Ite{S_1}{S}$,
and $S$ is not trivial.
In this case we distinguish two possibilities for the first element of the array, and we define 
$$N_{n+1}= (\Ite{S}{\True} \And \Ex_1 \Not S)\ \Or\  (\Ite{\Not S}{\True} \And \Ex_2 \Not S)$$

\end{enumerate}


Observe that in the first three cases we can express negation using the operators $\Ex_1 S$ and
$\Ex_n^M \True$ that where already present in Draft 06. In the fourth case, however, we need the operator
$\Ex_2  S$ that has been introduced in Version 2019/09.

We have examined a set of ca.\ 11,000 different schemas, which contain a total of 33,015 
instances of  $\Ite{S_1 \itdots S_n}{S}$.
Almost all of  those instances fall in cases 1 (97\%) and 2 (2,5\%), but we have 121 examples  of 3 (0,4\%), 
while case 4 covers seven cases. 
We found
only one schema that falls out of this classification, since it has $2$ item types and a non-trivial $S$ types.

While these four cases are sufficient in practice, we present here a general formula that is applicable to every
case.

Given $\Not\Ite{S_1 \itdots S_n}{S}$, let us divide any array that does not satisfy it in two parts:
the \emph{head} up to position $n$,
and the \emph{tail} (which may be empty) after position $n$.
The formula $N_{n+1}$ specifies that the tail is not empty and contains one element that violates $S$, that is, we have
some \emph{tail-non-$S$}'s.
We cannot directly express this, but we can reason by cases on the 
positions of the elements in the head that violate $S$ --- 
we call these elements the \emph{head-non-$S$}'s.
An array with some tail-non-$S$ can be described as an array that has $k$ \emph{head-non-$S$}'s and 
satisfies $\Ex_{k+1}\Not S$. Hence, the formula $N_{n+1}$ will enumerate all possible distributions of the head-non-$S$'s, 
and ask that one of these distributions holds, with $k$ head-non-$S$ and with $\Ex_{k+1}\Not S$, which
together imply
that one tail-non-$S$ exists.

In order to enumerate all distributions we consider, for the given $n$, the set of all \emph{bitmap}'s of length
$n$, where a bitmap is a function from $1..n$ to $\{0,1\}$, and, for a bitmap $bm$, we use
$\Sum(bm)$ for the number of its 1's, that is, for $\Sigma_{i \in 1..n} bm(i)$.

We define a function \NotIf(Bit,$S$)  such that
\begin{tabbing}
\NotIf(0,$S$) = $S$  \\
\NotIf(1,$S$) = $\Not S$
\end{tabbing}

Every bitmap $bm$ will correspond to a possible distribution of head-non-$S$'s, as follows that is described by the following schema:
$$
\Ite{\NotIf(bm(1),S)\itdots \NotIf(bm(n),S)}{\True} \And \Ex_{\Sum(bm)+1}\Not S
$$
The schema is satisfied by any array where the 1's of $bm$ indicate the positions of the head-non-$S$'s, and 
where at least $\Sum(bm)+1$ elements are non-$S$'s.
Hence any array $[J_1,\ldots,J_m]$ that satisfies that schema has $\Sum(bm)$ head-non-$S$'s
and some tail-non-$S$'s, and, vice versa, for every array $A$ that has some tail-non-$S$'s, there exists
a bitmap $bm$ such that $A$ satisfies the corresponding schema.

Hence, $N_{n+1}$ can be defined by the following disjunction with $2^n$ cases:
$$
\BigOr_{bm \in \SetTo{n}\To\Set{0,1}} \Ite{\NotIf(bm(1),S)\itdots 
\NotIf(bm(n),S)}{\True}  \And  R\\ 
$$
where
$$
R = \Ex_{\Sum(bm)+1}\Not S
$$

\comment{To sum up we have:
$$
\begin{array}{lllll}
\multicolumn{4}{l}{
\Not\Ite{S_1,\ldots,S_n}{S} = \TArr \And (N_1 \Or \ldots \Or N_n \Or N_{n+1})
}\\[\NL]
i\in\SetTo{n}&N_i & =& \Ite{\True_1\itdots\True_{i-1}\cdot\Not S_i}{\True} \And \Ex_i^{\Inf}\True \\[\NL]
&N_{n+1} &=& \BigOr_{bm \in \{0,1\}^n}(\ \Ite{\NotIf(bm(1),S)\itdots \NotIf(bm(n),S)}{\True} \\
& & & \qquad\qquad\qquad\And \Ex_{\Sum(bm)+1}^{\Inf}\Not S \ )
\end{array}
$$}

To sum up we have the  four formulas described in Figure \ref{fig:fourf}, where the last one subsumes the first three cases.
\begin{figure*}
\[
\begin{array}{lllll}
\Not(\Ite{}{S}) &=& \Type(\Arr) \And  & \Ex_1^{\Inf}\Not S \\[\NL]
\Not(\Ite{S_1 \itdots S_n}{\False}) &=&
\Type(\Arr) \And & (\BigOr_{i\in\SetTo{n}}(\Ite{\True_1\itdots\True_{i-1}\cdot\Not S_i}{\True} \And \Ex_i^{\Inf}\True) \ \ \Or  \Ex_{n+1}^{\Inf}\True) \\[\NL]
\Not(\Ite{S_1 \itdots S_n}{\True}) &=&
\Type(\Arr) \And &\BigOr_{i\in\SetTo{n}}(\Ite{\True_1\itdots\True_{i-1}\cdot\Not S_i}{\True} \And \Ex_i^{\Inf}\True)
\\[\NL]
\Not(\Ite{S_1 \itdots S_n}{S}) &=&
\Type(\Arr) \And & (\BigOr_{i\in\SetTo{n}}(\Ite{\True_1\itdots\True_{i-1}\cdot\Not S_i}{\True} \And \Ex_i^{\Inf}\True) \ \ \Or \\
& & & \ \ \BigOr_{bm \in 2^n}(  \Ite{\NotIf(bm(1),S)\itdots \NotIf(bm(n),S)}{\True} \And \Ex_{\Sum(bm)+1}^{\Inf}\Not S \ )
\end{array}
\]
\caption{Items negation formulas.}
\label{fig:fourf}
\end{figure*}

\comment{
\begin{remark}
It is easy to see that this formula reduces to the four versions of $N_{n+1}$ that we defined above: 
in cases 1  ($n=0$) and 4 ($n=1$) we respectively have  $\Sum(bm)=0$ and $\Sum(bm)\in\{1,0\}$.

In case 3 ($S=\True$ or $S \supseteq S_i$)),
observe that, if a bitmap includes a 1 in position $i$, then the corresponding disjunct inside 
$N_{n+1}$ 
requires the presence of a non-$S$ at position $i$, but, since $\Not S \subseteq \Not S_i$, this case is already 
covered by $N_i$, hence the only non-redundant disjunct is the one for the bitmap 0...0, and, also in that case, the 
$\Ite{}{}$ part of  the remaining disjunct of $N_{n+1}$:
$$
 \Ite{\NotIf(0,S)\itdots \NotIf(0,S)}{\True} \And \Ex_{\Sum(bm)+1}^{\Inf}\Not S
$$
is implied by $N_0,\ldots,N_n$, hence we rest with $\Ex_{1}^{\Inf}\Not S$.

In case 2 ($S=\False$) we only consider the 
bitmap 1...1, since all the other cases contain a $\False$ in one of the first $n$ positions hence
cannot be satisfied by an array whose length is greater than $n$,
where $\Ite{\NotIf(bm(1),S)\itdots \NotIf(bm(n),S)}{\True}$ is equivalente to $\True$,
and ignore all the other cases, where the term is equivalent to $\False$.
\end{remark}
}
\comment{
While this exponential explotion looks bad, in practice it is never necessary. The vast majority of 
uses of \xaddIts\ presents a value of $\xfalse$, where the only satisfiable addend of $N_{n+1}$ 
is the one that corresponds to the 1...1 bit map, that is:

$$
\Ite{\True_1,\ldots,\True_n}{\True} \And \Ex_{n+1}\True 
$$

This can be further simplified to $\Ex_{n+1}\True$ since the first part is always true.

The second common case is when \xaddIts\ presents a value of $\xtrue$. In this case $N_{n+1}$ is
equivalent to $\False$   because of
the $\Ex_{\Sum(bm)+1}\Not S$ that is found in any addend, which corresponds to the fact that the 
\xaddIts : \xtrue\ assertion is trivial hence cannot be violated.

This case can be generalized to the case $\Ite{[S_1,\ldots,S_n]}{S'}$ where $S'$ includes (in its semantics) all of
$S_1,\ldots,S_n$. 
In this case, $N_{n+1}$ can just be $\Ex_{1}^{\Inf}(\Not S_{n+1})$:
to prove that $N_{n+1} \Implies \Not(\Ite{}{})$ observe that, since $\Not S_{n+1}$ is disjoint from
 $S_1,\ldots,S_n$, any element with that type, in any position, would violate the type.
 To prove that $\Not \Ite{}{} \And \Not(N_1) \And \ldots \And \Not(N_n) \Implies N_{n+1}$,
every array that violates $\Ite{}{}$ but has correct values in all the first $n$ positions must violate
$S_{n+1}$ in a position that is greater than $n$.

Out of 988 uses we examined, we only found six that are not captured by the previous three patterns, but all
the three of them have an array of length one, hence they only need two bitmaps for the 
}

\comment{
\subsection{Not-pushing}\label{sec:notpushing}

Not-pushing is the process of pushing negation through the boolean operators down to the 
typed operators.

These are the not-pushing rules. We also present a rule to push regation across
not-completed variable definitions, but this is never needed.  
$$\begin{array}{llll}
\Not (S_1 \And S_2) &=& (\Not S_1) \Or (\Not S_2) \\[\NL]
\Not (S_1 \Or S_2) &=& (\Not S_1) \And (\Not S_2) \\[\NL]
\Not \True &=& \False \\[\NL]
\Not \False &=& \True \\[\NL]
\Not \XOr( S_1,\ldots,S_n; \Not S_1,\ldots,\Not S_n)&=&
\NotXOr(\Not S_1,\ldots,\Not S_n; S_1,\ldots,S_n)\\[\NL]
\Not \NotXOr(S_1,\ldots,S_n;\Not S_1,\ldots,\Not S_n)&=&
\XOr(\Not S_1,\ldots,\Not S_n; S_1,\ldots,S_n) \\[\NL]
\Not (\Not S) &=& S \\[\NL]
  \!\!\!  \begin{array}{lllll}
      \Not ( \ADef{x_0}{S_0} , \ldots, \\
  \quad            \Def{x_n}{S_n}\\
   \quad           \Def{not\_x_0}{S_{n+1}} , \ldots, \\
     \quad     \Def{not\_x_n}{S_{2n}}\ \ \ \ \ \ )
   \end{array}
& = &
 \!\!\!  \begin{array}{lllll}
              \Def{x_0}{S_0} , \ldots, \\
              \Def{x_n}{S_n}\\
              \ADef{not\_x_0}{S_{n+1}} , \ldots, \\
              \Def{not\_x_n}{S_{2n}}
  \end{array} 
\end{array}
$$
}

\subsection{Not rewriting}

We show now how to push negation down any algebraic expression. 


Not-elimination may generate bounds that are trivial or unsatisfiable,
hence we should apply bound-normalization rules, such as the following ones.
We report the case for $\Pro$. Similar rules exist for $\Bet$, $\XBet$,  $\Ex$. 

$$\begin{array}{llll}
\Pro_0^{\Inf} & = & \True \\[\NL]
\Pro_n^{m} & = & \TObj \Implies \False & \text{if } n > m \\[\NL]
\Pro_{\Inf}^{\Inf} & = & \TObj \Implies \False & \\
\end{array}
$$

\comment{
Not elimination uses the six negative operators listed in Section \ref{sec:full}.
The operator $\NotPat(r)$ is actually equivalent to $\Pat(not\ r)$ where $not\ r$ is the pattern that matches everything that is not matched by $r$.
As discussed in section \ref{sec:names} 
applying the same pattern-invertion operation, it is possible to translate both $\Nam(S)$ and $\ExNam(S)$ in $r :: \False$, with a pattern $r$ that depends on $S$.
In the same way, we can translate 
$\APReq((r_1,\ldots,r_n):S) $ into $\PReq(\NotP(r_1|\ldots|r_n):S)$.
Hence, if we consider the possibility of pattern rewriting, the only negative
operators that are really needed for not-elimination are: 
$\NotMof(n)$, $\NotUni$, and $\PReq$. Every other operator can be complemented 
using another positive JSON schema operator.

\GG{Suggestion for the implementors: please consider the possibility to 
implement $\Type[T_1,\ldots,T_n]$,
so that the complent of $\Type(T)$ does not really need a disjunction}
}

Not elimination is defined as follows. We do not define the cases for the negative operators
($\NotMof$ etc.) since they follow immediately from their definitions.
$$\begin{array}{llll}
\Not (S_1 \And S_2) &=& (\Not S_1) \Or (\Not S_2) \\[\NL]
\Not (\Not S) &=& S \\[\NL]
\Not(\Type(T)) & = & \Or( \Type(T') \M T' \neq T ) \\[\NL]
\Not(\Pat(r)) &=& \Type(\Str) \And \NotPat(r) \\[\NL]
\Not(\Bet_{m}^{M}) & = &
\Type(\Num) \And (\XBet_{-\Inf}^{m} \Or \XBet_{M}^{\Inf}) \\[\NL]
\Not(\IBT(\xfalse)) &=& \TBool \And\IBT(\xtrue) \\[\NL]
\Not(\IBT(\xtrue)) &=& \TBool \And \IBT(\xfalse) \\[\NL]
\Not(\Mof(n)) & = &
\Type(\Num) \And \NotMof(n) \\[\NL]
\Not(r:S) & = & \Type(\Obj) \ \And \PReq(r:\Not S)
\\[\NL]
\Not(\Pro_{i}^{j}) & = &
\Type(\Obj) \And (\Pro_{0}^{i-1} \Or \Pro_{j+1}^{\Inf}) \\[\NL]
\Not(\Ite{S_1 \itdots S_n}{S_{n+1}}) &=&
\text{See Section \ref{sec:notitem}}\\[\NL]
\Not(\Ex_{i}^{j}S) & = &
\Type(\Arr) \And (\Ex_{0}^{i-1}S \Or \Ex_{j+1}^{\Inf}S) \\[\NL]
\Not(\Uni) & = &
\Type(\Arr) \And \NotUni\\[\NL]
\Not(\Ref{x}) &=& \NotVar{x}\\[\NL]
  \!\!\!  \begin{array}{lllll}
      \Not ( \ADef{x_0}{S_0} , \ldots, \\
  \quad            \Def{x_n}{S_n}\\
   \quad           \Def{not\_x_0}{S_{n+1}} , \ldots, \\
     \quad     \Def{not\_x_n}{S_{2n}}\ \ \ \ \ \ )
   \end{array}
& = &
 \!\!\!  \begin{array}{lllll}
              \Def{x_0}{S_0} , \ldots, \\
              \Def{x_n}{S_n}\\
              \ADef{not\_x_0}{S_{n+1}} , \ldots, \\
              \Def{not\_x_n}{S_{2n}}
  \end{array} 
\end{array}
$$

\hide{
Negation of $\Props$ requires the use of negative operators when we have 
non-trivial patterns or a schema for the additional properties that is not $\True$.
The vast majority of objects schema do not make use of $\xpattProps$ and impose
no constraint for the additional properties, and in this case we can have full
not-elimination, with no need of the special negative assertions, as follows. 
}


\begin{example}
For example, assume the following definition.
$$
\begin{array}{lllll}
\ADef{x}{\{ a : \NR{x}\}}
\end{array}
$$
where $\NR{x}$ is the complement variable.
This is the effect of completion.
$$
\begin{array}{lllll}
\ADef{x}{\{ a : \NR{x}\}},
\Def{not\_x}{\Not \{  a : \NR{x}\}} 
\end{array}
$$

And this is how not-elimination may now proceed.

\hide{Consider the following schema, and try to understand its meaning:
$\ADef{x}{\{ a : \NR{x}\}}$

This may become clearer after performing first not-elimination, as follows.}
$$
\begin{array}{lllll}
\ADef{x}{( a : \NR{x})},
&\!\!\Def{not\_x}{\Not (  a : \NR{x})}  \ \ \To \\[0.8ex]
\ADef{x}{{( a :\Ref{not\_x})}},
&\!\!\Def{not\_x}{( \Type(\Obj)\And  \Req(a)\And  a :  \Not \NR{x})} \ \ \To \\[0.8ex]
\ADef{x}{{( a :\Ref{not\_x})}},
&\!\!\Def{not\_x}{( \Type(\Obj)\And  \Req(a)\And  a :  \Ref{x})} \\[0.8ex]
\end{array}
$$

where $\Req(a)$ denotes the fact that $a$ is \emph{required}, i.e.
$$\Req(a) = \TObj \Implies \Not ( \keykey{a} : \False) $$

We can now substitute \key{not\_x} with its definition.
The definition that we get is not much clearer: \emph{if} the value is an object with an \emph{a}
member, then the value of that member \emph{must } be an object with an \emph{a} member,
whose value satisfies the same specification.
$$
\begin{array}{lllll}
\ADef{x}{{( a : (\Type(\Obj) \And \Req(a) \And a :  \Ref{x}) )}}
\end{array}
$$

Some examples of values that match that schema:
$$
1, \{ a : \{ a : 1 \} \} , \{ a : \{ a : \{ a : \{ a : 1 \} \}  \} \} 
$$
\end{example}


\section{Witness generation}\label{sec:witness}

\subsection{The structure of the algorithm}

In order to prove satisfiability, or emptiness, of a schema, we try and generate a 
witness for the schema. We examine all the possible ways to generate the value hence,
if generation is not possible, the schema is not satisfiable.

The basic idea is as follows.
Assume, by induction, that you have an algorithm to generate a witness for any assertion
$S$ of size up to $n$. In order to generate a witness for an ITE of size $n+1$ such as
$\PReq(r:S)$ one will generate a witness for $S$ and use it to build an object with 
a member that matches $r$ whose value is that witness, and the same approach can be
followed for the other ITEs. For the boolean operator
$S_1 \Or S_2$, we recursively generate witnesses of $S_1$ and of $S_2$. Negation and conjunction
are a problem: there is no way to generate a witness for $\Not S$ starting from a witness 
for $S$, and, given a witness for $S_1$, if this is not a witness for $S_1 \And S_2$, we may
need to try infinitely many others before finding one that satisfies $S_2$.
Hence, we first eliminate $\Not$ using not elimination, then we bring all definitions 
of variables into DNF so that conjunctions are confined to the leaves of the syntax tree,
and finally we make conjuction harmless by a technique that is called \emph{canonicalization},
which is based on the combination of all ITEs having to do with the same type, and which is
presented below. After ITEs and boolean operators, we are left with recursive
variables. We deal with them by adopting a bottom-up iterative evaluation that mimics the
fixpoint semantics. 
We are now going to transform this idea into an algorithm.

For the sake of presentation, we will only present our approach by focusing on schemas that  do not feature  $\Uni$ or $\NotUni$. 

\comment{
In order to deal with the $\Uni$ constraint of array types, we do not content ourselves with
an algorithm that shows one witness for any non-empty type, but we will produce an algorithm
that produces a \emph{dense} enumeration of each non-empty type, where we call \emph{dense}
and enumeration that is complete on all types whose domain is finite and that returns an unbounded
number of witnesses for those types whose domain is infinite.
We will often express the algorithm in terms of non-determistic choice in a given set.
The interpretation is that the algorithm will enumerate all the elements of the set if it is
finite, and an unbounded subset if it is infinite.}

The algorithm consists in six steps.

\begin{enumerate}
\item Translation from {\jsonsch} to the core algebra and not-elimination.
\item Canonicalization: we split every conjunction into 
       a disjunction of 
       \emph{typed groups}, 
        where a typed group is a conjunction of typed assertions that contains one 
        $\Type(T)$ assertion, and where $T$ is the type of all the ITEs in the group.
   \item Variable normalization: we rewrite the definitions of all variables so that
       every variable only appears as an $S$ argument of a typed operator, as in
       $\PReq(a : x)$, and all $S$ arguments of typed operator are variables.
        In this way, no boolean operator has a variable as argument.
   \item Reduction to DNF: we transform each boolean expression into a Disjunctive Normal Form,
   that is, into a disjunction of typed groups.
 \item Object and array preparation: we rewrite object and array groups into a form
    that will simplify recursive generation of witnesses. In a sense, we
   precompute all
        intersection types that we may need during the next phase.
\item Recursive generation: we start from a state where the semantics of every variable is   
     unknown,
      and we repeatedly try and generate a witness for each variable, until a fixpoint is
      reached.   
\end{enumerate}

Before presenting the above steps, we introduce and-merging, that is not a step of the algorithm, and is not mandatory, 
but is an optimization step that reduces the size of any conjunction
by combining related operators, and that can be reapplied at any time
a new conjunction is generated when the above tasks are performed. 
we describe the six steps of the algorithm.

\subsubsection{And-merging}This operation consists of 
rewriting the conjunction of two expressions into on equivalent one that is  simpler. For instance, the schema
$\Bet_{m_1}^{M_1}\ \And\ \Bet_{m_2}^{M_2}$ is equivalent to
$\Bet_{\max(m_1,m_2)}^{\min(M_1,M_2)}$, while 
$\TInt \And \Pro_0^4$ is equivalent to $\TInt$.

For optimization purposes, the biggest gain is reached when we discover inconsistencies that allow an expression to
be rewritten as $\False$, which can be then propagated to a surrounding conjunctive context.

And-merging is specified as a set of rules $S_1, S_2 \To S_3$ that specify that, whenever 
$S_1$ and $S_2$ appear in any conjunction of conjunctions, 
in any positions, they are rewritten as $S_3$. When $S_3$ is $\False$, the entire
conjunction is rewritten as $\False$.

The algebra enjoys a huge set of and-merging rules. We focus here, 
for space reasons, on a subset of the cases we really deal with (see \cite{long} for more details).

\setlength{\NL}{0ex}

The following three rules are the most important ones.
After not-elimination, the only typed operators left are some ITEs 
and unary type assertion $\Type(T)$.
The first two rules say that only one $\Type$ operator survives 
and-merging, and the third rules specifies that any ITE that 
limits a type that is different from $T$ is redundant in
presence of $\Type(T)$ since every value of type $T$ 
trivially satisfies any ITE whose type is different from $T$.
$$\begin{array}{llllll}
\Type(T), \Type(T) &\To & \Type(T) & \\[\NL]
\Type(T), \Type(T') &\To & \False & T \cap T' = \Emptyset \\[\NL]
\Type(T), S &\To & \Type(T) & S \in \key{ITE}(T'), \ T' \cap T = \Emptyset\\[\NL]
\end{array}
$$

\comment{
$\Nam$ and $\ExNam$ are also and-merged together.
The condition $S_1 \subseteq S_2$ means that the rule can be applied whenever
the rewriting tool is able to prove this condition, which should include reflexivity, minimality of
$\False$, and maximality of $\True$.
After merging, we have one $\Nam$ at most, but possibly many $\ExNam$.
$$\begin{array}{llllll}
\Nam(S_1) \And \Nam(S_2) &\To & \Nam(S_1 \And S_2) \\
\Nam(\False) & \To & \Pro_0^0  \\ 
\Nam(S_1) \And \ExNam(S_2) &\To & \Nam(S_1) \And \ExNam(S_2 \And S_1) \\
\ExNam(\False) & \To & \TObj \Implies \False \\
\ExNam(S_1) \And \ExNam(S_2) &\To & \ExNam(S_1) & S_1 \subseteq S_2 \\ 
\end{array}
$$

The other object operators are and-merged in a very basic way.
In the rules below, $K$ is either a set of names or a set of patterns
where each pattern matches a single name, $R$ a set of patterns,
$P$ is a set of 
pattern-schema pairs, = is semantic equality, that is, denoting the same set of 
name for two patterns or the same set of instances for two assertions,
and $r \SubP r'$ means that every name matched by $r$ is also matched by $r'$.
Since these are just optimizations, the tool will apply these rules when equality
or inclusion can be proved without excessive efforts.
After and-merging, we have only one $\Req$ and one $\PReq$, 
but we could have many 
$\Props(r : S \And  S', r_1 : S_1,\ldots,r_n : S_n;S'') $
statements. Their complete and-merging is part of object type
preparation (Section \ref{sec:objprep}).
$$\begin{array}{llllll}
\Req(K_1) , \Req(K_2)  &\To & \Req(K_1 \cup K_2)  \\[\NL]
\PReq(P_1), \PReq(P_2) &\To & \PReq(P_1, P_2) \\[\NL]
\!\!\!\begin{array}{lll}
\APReq(R : S), \\[\NL]
\qquad\APReq(R' : S')\\[\NL]
\end{array} &\To &
\APReq(R : S \And S') & R = R'\\[\NL]
\Props(P_1;\True), \Props(P_2;\True)
& \To & \Props(P_1, P_2;\True)\\[\NL]
\Props(r : S, r' : S', P ; S'') &\To &
\Props(r : S \And  S', P;S'')  & r = r' \\[\NL]
\Props(r : S, r' : S', P ; S'') &\To &
\Props(r|r' : S, P;S'')  & S = S' \\[\NL]
\PReq(r : S, r' : S', P ; S'') &\To &
\PReq(r : S, P;S'')  & r \SubP r'  \ \And \ S \subseteq S' \\[\NL]
\end{array}
$$
}

Here we present the object and-merging rules.
The symbol $S \FastSub S'$ indicates any easy-to-prove relation that implies
inclusion, such that the relation where the two expressions are equal or
where $S'=\True$ or $S=\False$. The exact relation is irrelevant, as far as
it implies inclusion, since and-merging is just an optional fast optimization.
In the same way, equality in the second line may be just syntactic equality 
enriched with any easy-to-compute equivalence rule. 
The pattern $r|r'$ matches the union of $r$ language and $r'$ language.
We use $r \SubP r'$ to denote regular expression inclusion, which is 
decidable.
$$\begin{array}{llllll}
\CProps{p: S_1}, \CProps{p: S_2}
& \To & \CProps{p: S_1 \And S_2} &\\[\NL]
\CProps{p : S},\CProps{p' : S'} \qquad\qquad &\To & \CProps{p|p' : S}  & S = S' \\[\NL]
\multicolumn{3}{l}{\PReq(r : S), \PReq(r' : S')}\\
 &\To &
\PReq(r : S) & r \SubP r'  \ \And \ S \FastSub S' \\[\NL]
\end{array}
$$
We finally present the array operators, apart from $\Ex_i^j  S$.
We show that all instances of $\Ite{}{}$ can be always merged into one.

$$\begin{array}{llllll}
\Ite{S_1, \itdots, S_n}{S_{n+1}},
\Ite{S'_1, \itdots, S'_n}{S'_{n+1}} \To \\[\NL]
  \quad   \Ite{S_1 \And S'_1, \itdots, S_n \And S'_n}{S_{n+1}\And S'_{n+1}} \\[2mm]
\Ite{S_1, \itdots, S_n}{S_{n+1}},\\[\NL]
\quad \Ite{S'_1, \itdots, S'_m}{S'_{m+1}} \To  \Ite{S}{S_{n+1}\And S'_{m+1}} \\[\NL]
\text{with } m > n  \\ [\NL]
\text{and } S=S_1 \And S'_1 \itdots S_n \And S'_n \cdot S_{n+1}\And S'_{n+1}\itdots S_{n+1}\And S'_{m} \\ [2mm]

\Uni, \Uni \To  \Uni \\[\NL]
\NotUni, \NotUni \To  \NotUni \\[\NL]
\Uni, \NotUni \To  \TArr \Implies \False \\[\NL]
\end{array}
$$

\noindent Rules for other and simpler cases are  in the full report~\cite{long}.  

\subsection{Translation to JSON Schema and Not-elimination}

We first translate {\jsonsch} to the core algebra, but we keep the $\Ite{}{}$ notation,
 rather than the $i \D j : S$ notation of the core,
since it is notationally more convenient.
Not-elimination is then performed as in Section \ref{sec:notelimination}.

\comment{eliminate some operators.
Specifically, we eliminate $\Con(J)$ and $\Enu(J_1,\ldots,J_n)$,
we translate $n$-ary $\Type(T_1,\ldots,T_n)$ as disjunctin of unary $\Type(T)$,
we translate $\Type(\Int)$ into $\TNum and \Mof(1)$, and we translate all object operators
into the not-eliminated core algebra operators. We also translate }

\comment{As a preliminary step, we transform any $\Enu$ that is not type-homogeneous into
a disjunction of its type-homogenous components so that in the rest of the paper we can
assume that every $\Enu$ is type-homogeneous.
We also transform $\Con(J)$ into $\Enu(J)$.}

\hide{
\begin{remark}
As an optimization, for the operator $\XOr(S_1,\ldots,S_n)$, in order to reduce the size explosion
during DNF transformation , we do not immediately rewrite it but, in this phase, we just
rewrite it as $\XOrVar(S_1,\ldots,S_n;\Not S_1,\ldots,\Not S_n)$,
in order to make negation explicit, and we define a dual operator
$\NotXOr(S_1,\ldots,S_n;\Not S_1,\ldots,\Not S_n)$, defined as 
$\Not \XOrVar(\Not S_1,\ldots,\Not S_n; S_1,\ldots,S_n)$, where
$\NotXOr(S_1,\ldots,S_n)$ means:
either all schemas are satisfied, or at least two schemas are violated.
We perform not elimination keeping both $\XOrVar$ and $\NotXOr$, and we transform them at the end
of not-elimination, as follows:
$$
\begin{array}{llll}
\XOrVar(S_1,\ldots,S_n; S'_1,\ldots,S'_n) & = &
\BigOr_{1 \leq i \leq n} \And({S'_1},\ldots, {S'_{i-1}},{S_1}, {S'_{i+1}},\ldots, {S'_n})\\[\NL]
\NotXOr(S_1,\ldots,S_n; S'_1,\ldots,S'_n) & = &\And({S_1},\ldots,{S_n}) \Or \BigOr_{1 \leq i < j \leq n}({S'_i} \And {S'_j})
\end{array}
$$

Of course, if we can prove  that all the arguments of $\XOr(S_1,\ldots,S_n)$ are mutually incompatible, which is often the case,
then we can just rewrite $\XOr$ as $\Or$.
\end{remark}
}

\comment{
As a second optimization, we do not transform $\Enu$ and $\Con$ as specified in Section
\ref{sec:notelimination}, but we use a negative operator $\NotEnu(M)$ defined
as $\Not (\Enu(M))$
(to be better expained), which is more convenient for witness generation.}

\comment{
\begin{remark}
Otherwise, we should first rewrite it using $\DefKey$ as follows:
$\ADef{\_}{\XOr(\Ref{x}_1,\ldots,\Ref{x}_n)}, 
\Def{x_1}{S_1},\ldots, \Def{x_n}{S_n}$, which will allow us to compute the negation
of each $S_i$ just once.
At this point, we can rewrite
$\XOr(\Ref{x}_1,\ldots,\Ref{x}_n)$
in DNF as
$$\Or_{1 \leq i \leq n} \And(\NR{x}_1,\ldots,\NR{x}_{i-1},\Ref{x}_1, \NR{x}_{i+1},\ldots, \Not     \Ref{x}_n)$$
Similarly, we can rewrite $\NotXOr(S_1,\ldots,S_n)$
in DNF as
$$\And(\Ref{x}_1,\ldots,\Ref{x}_n) \Or \Or_{1 \leq i < j \leq n}(\NR{x}_i \And \NR{x}_j) $$
Both forms need $n^2$ atoms in their expression.
\end{remark}
}

\subsection{Canonicalization}


Canonicalization is a process defined along the lines of \cite{habib2019type}. We rely on the the new notation
$\{ S_1,\ldots S_n \}$, which we call a \emph{group}, for representing  the $n$-ary conjunction of schemas, 
and we define \emph{canonicalization} of a not-eliminated expression, 
which may include both binary conjunctions and group conjunctions,  
as the following process.


\begin{enumerate}
\item We first flatten any tree of nested conjunctions, of both forms, into a single group. 
\item For every group  $\{T_1,\ldots,T_n, S_1,\ldots, S_m \}$, where $T_1,\ldots,T_n$ 
are the typed assertions and
$S_1,\ldots, S_m$ the boolean, definition, or variable assertions, we rewrite it as 
$\{T_1,\ldots,T_n\} \And S_1 \And \ldots \And  S_m$. 
Thanks to step (1), no $S_i$ is a conjunction.
\item For every group  $\{ G \}$ where at least one $\Type(T)$ assertion is present, we apply
and-merging until the group only contains one $\Type(T)$  and a set of ITEs with a compatible
type, or collapses to $\False$. 
\item For every group  $\{ G \}$ where no $\Type(T)$  is present, we rewrite it as the 
\emph{disjunction} of six groups, each one starting with a different $\Type(T)$ assertion, one for each 
core algebra type, and continuing with the subset of $G$ whose type is $T$.
The group is formed even if this subset is empty.
\end{enumerate}

As an example,  the following expression:
$$  (\Mof(3) \And \Len_0^5) \Or (\{ \TNum , \Mof(2) , \Ref{x} \And \Pat(a*) \} ) $$
is rewritten as follows.
$$\begin{array}{lll}
(1) & \{ \Mof(3), \Len_0^5\}  \Or \{ \TNum , \Mof(2), \Ref{x}, \Pat(a*)  \} \\[\NL]
(2) & \{ \Mof(3), \Len_0^5\}  \Or ( \{ \TNum , \Mof(2),  \Pat(a*) \} \And \Ref{x} ) \\[\NL]
(3) & \{ \Mof(3), \Len_0^5\}  \Or ( \{ \TNum , \Mof(2) \} \And \Ref{x} ) \\[\NL]
(4) & \{ \TNull \} \Or \{ \TBool \} \Or \{ \TNum, \Mof(3) \} \\[\NL]
      & \Or \{ \TStr, \Len_0^5 \}  \Or \{ \TArr \} \Or \{ \TObj \}  \\[\NL]
      & \Or ( \{ \TNum , \Mof(2) \} \And \Ref{x} )
\end{array}$$

We say that a group that contains $\Type(T)$ and a set of ITEs of type $T$ is a \emph{typed group} of type $T$.
At the end of the canonicalization phase, any expression has been rewritten as a boolean combination of variables, 
definitions, and typed groups. 
{
For reasons of space, hereafter we will abbreviate a type assertion $\TNum$ in a
group with $\Nu$, and similarly for the other types, 
so that we use $\Nl, \Bo, \Nu, \St, \Ob, \Ar$ to indicate the six core types. 
We will also use concatenation to indicate disjunction, so that
$\TG{\Nl\Bo\Nu} = \TG{\Nl} \Or \TG{\Bo} \Or \TG{\Nu}$, and we use $\overline{X}$ to indicate complement
with respect to $\Nl, \Bo, \Nu, \St, \Ob, \Ar$, so that $\TG{\overline{\Nl}}$ is the same as $\TG{\Bo\Nu\St\Ob\Ar}$.
In this way, the expression above can be written as follows. 

$$\begin{array}{lll}
(4) & \{ \TNull \} \Or \{ \TBool \} \Or \{ \TNum, \Mof(3) \} \\[\NL]
      & \Or \{ \TStr, \Len_0^5 \}  \Or \{ \TArr \} \Or \{ \TObj \}   \\[\NL]
      & \Or ( \{ \TNum , \Mof(2) \} \And \Ref{x} ) \\[2\NL]
=  & \TG{\overline{\Nu\St}} \Or \{\Nu, \Mof(3) \} \Or \{ \St, \Len_0^5 \} \\[\NL]
     &\Or
        ( \{ \Nu , \Mof(2) \} \And \Ref{x} )
\end{array}$$
}

\comment{
\subsection{The WG algebra}

$$
\begin{array}{llll}
S & ::= & A \M \IBT(b)\M \Pat(r) \M \Bet_{m}^{M}  \M \Mof(n) \M \NotMof(n) 
            \M \Type(T)   \\[\NL]
&&  \M \Pro_{i}^{j} \M \key{r} : A  \M \OPReq(r : A)  \\[\NL]
&&  \M i  : A \M i \D \Inf : A \M \Ex_{i}^{j}A \M  \Uni \M \NotUni \\[\NL]
&& 
 \M \ S_1 \And S_2   \M  \ADef{x_1}{S_1} , \ldots, \Def{x_n}{S_n} \M \Ref{x} \\[\NL]
A & ::= & \True \M \False \M \Ref{x} \\[\NL]
T & ::= & \Arr \M \Obj \M \Null \M \Bool \M \Str  \M \Num \M \Int \\[\NL]
r & ::= &  \text{\em {\jsonsch} regular expression}  \\[\NL]
b & ::= & \xtrue \M \xfalse 
\end{array}
$$

The witness generation algorithm uses the first three steps to transform full-algebra
expressions into equivalent canonical expressione written in its own WG-algebra,
and then reduces 
 until these expressions are ready to 
act as witness generators 
}

\subsection{Variable normalization}\label{sec:varnorm}

Variable normalization is used to reach a form where no boolean operator has a 
variable as an argument, and all non-boolean schema operators only have variables as arugment.
This will be crucial for the  witness generation phase (Section  \ref{sec:recgen}). 
It proceeds in two steps, \emph{separation}
and \emph{expansion}.

In the \emph{separation} phase, for every typed operator that has a subschema $S$ in its syntax, 
such as $\Ex_i^j S$, when $S$ is not a variable
we add to the global definition a new variable definition $\Def{x}{S}$, and we substitute 
$S$ with $\Ref{x}$.\footnote{In the implementation we make an exception for $\True$
and $\False$, which can appear wherever a variable appears.}
For every variable $\Def{x}{S}$ that we define, we must also define its complement
 $\Def{not\_x}{\Not S}$, and perform not-elimination and canonicalization on $\Not S$. 

In the \emph{expansion} phase, 
for any clause  $\Def{x}{S}$, we substitute any unguarded
occurrence of any variable in $S$, where an occurrence is guarded if it occurs below a typed operator,
with its definition. This process is guaranteed to stop since we do not allow unguarded
cyclic definition.

For example, consider the following definitions.
$$\begin{array}{llll}
\ADef{x}{(\{\Ar , \Ite{\Ref{x}\And\Ref{y}}{\True}\}\And \Ref{y}) \Or \TBo} \\[\NL]
\Def{y}{\{\Ar , \Ite{\TNu \Or \Not \Ref{x}}{\True}\} }\\[\NL]
\end{array}$$

By not-elimination and canonicalization, we obtain the following schema.
$$\begin{array}{llll}
\ADef{x}{(\{\Ar , \Ite{\Ref{x}\And\Ref{y}}{\True}\}\And \Ref{y}) \Or \TBo} \\[\NL]
\Def{y}{\{\Ar , \Ite{\TNu \Or \Ref{not\_x} }{\True}\} }\\[\NL]
\Def{not\_x}{(\{\Ar , \Ite{\Ref{not\_x}\Or\Ref{not\_y}}{\True}, \Ex_1^{\Inf}\True \}
                   \Or  \TNArr \Or \Ref{not\_y}) } \\[\NL]
                     \qquad  \qquad  \qquad \And \TNBool  \\[\NL]
\Def{not\_y}{\{\Ar , \Ite{\TNNum\And \Ref{x}}{\True}, \Ex_1^{\Inf}\True\} \Or \TNArr }\\[\NL]
\end{array}$$

The typed operator $\Ite{}{}$ is applied to a non-variable subschema $\Ref{x}\And\Ref{y}$ in 
$ \Ite{\Ref{x}\And\Ref{y}}{\True}$ in the first line, and similarly in the other three lines.
Hence, during the separation phase we define four new variables in order to separate
the non-variable arguments from their guarded operators. 
We should define four more variables in order to complete the generated equation system, 
but this is not necessary
in this case, since, for example, $\Ref{itnx}$ corresponds already to the negation of $\Ref{itx}$.
Separation produces the following schema.
$$\begin{array}{llll}
\ADef{x}{(\{\Ar , \Ite{\Ref{itx}}{\True}\}\And \Ref{y}) \Or \TBo} \\[\NL]
\Def{itx}{\Ref{x}\And\Ref{y}} \\[\NL]
\Def{y}{\{\Ar , \Ite{\Ref{ity}}{\True}\} }\\[\NL]
\Def{ity}{\TNu \Or \Ref{not\_x}} \\[\NL]
\Def{not\_x}{(\{\Ar , \Ite{itnx}{\True}, \Ex_1^{\Inf}\True \}
                   \Or  \TNArr \Or \Ref{not\_y}) \And \TNBool}  \\[\NL]
\Def{itnx}{\Ref{not\_x}\Or\Ref{not\_y}} \\[\NL]
\Def{not\_y}{(\{\Ar , \Ite{itny}{\True}, \Ex_1^{\Inf}\True\} \Or \TNArr)}\\[\NL]
\Def{itny}{\TNNum \And \Ref{x}} \\[\NL]
\end{array}$$

Now, we must expand all the unguarded variables. 
The unguarded variables are those that are underlined below.
$$\begin{array}{llll}
\ADef{x}{(\{\Ar , \Ite{\Ref{itx}}{\True}\}\And \underline{y}) \Or \TBo} \\[\NL]
\Def{itx}{\underline{x} \And \underline{y}} \\[\NL]
\Def{y}{\{\Ar , \Ite{\Ref{ity}}{\True}\} }\\[\NL]
\Def{ity}{\TNu \Or \underline{not\_x}} \\[\NL]
\Def{not\_x}{(\{\Ar , \Ite{itnx}{\True}, \Ex_1^{\Inf}\True \}
                   \Or  \TNArr \Or \underline{not\_y} ) \And \TNBool}  \\[\NL]
\Def{itnx}{\underline{not\_x} \Or \underline{not\_y}} \\[\NL]
\Def{not\_y}{\{\Ar , \Ite{itny}{\True}, \Ex_1^{\Inf}\True\} \Or \TNArr}\\[\NL]
\Def{itny}{\TNNum \And \underline{x}} \\[\NL]
\end{array}$$

One can observe that, because of mutual recursion, a process of iterated
variable expansion may never stop. However, by the assumption of guarded recursion,
the dependencies between
the unguarded occurrences are not cyclic. In this case, the longest dependency path is
\Ref{itx} depends of $\Ref{x}$ that depends on $\Ref{y}$, and similarly for 
\Ref{ity}, \Ref{itnx} and \Ref{itny}.


At this point, variable expansions  steps  produce the following set.
$$\begin{array}{llll}
\ADef{x}{}&(\{\Ar , \Ite{\Ref{itx}}{\True}\}\And \{\Ar , \Ite{\Ref{ity}}{\True}\})
 \Or \TBo \\[\NL]
\Def{itx}{}&((\{\Ar , \Ite{\Ref{itx}}{\True}\}\And \{\Ar , \Ite{\Ref{ity}}{\True}\}) \Or   \\[\NL]
& \qquad  \TBo)  \And \{\Ar , \Ite{\Ref{ity}}{\True}\} \\[\NL]
\Def{y}{}&\{\Ar , \Ite{\Ref{ity}}{\True}\} \\[\NL]
\Def{ity}{}&\TNu \Or ((\{\Ar , \Ite{itnx}{\True}, \Ex_1^{\Inf}\True \}  \Or  \TNArr \\[\NL]
                  &  \Or \{\Ar , \Ite{itny}{\True}, \Ex_1^{\Inf}\True\} \Or \TNArr ) \And \TNBool) \\[\NL]
\Def{not\_x}{}&(\{\Ar , \Ite{itnx}{\True}, \Ex_1^{\Inf}\True \}
                   \Or  \TNArr \Or \\[\NL] 
                   & \{\Ar , \Ite{itny}{\True}, \Ex_1^{\Inf}\True\} \Or \TNArr ) 
                   \\[\NL]
                   & \And \TNBool  \\[\NL]
\Def{itnx}{}&((\{\Ar , \Ite{itnx}{\True}, \Ex_1^{\Inf}\True \}
                   \Or  \TNArr  \\[\NL] 
                   & \Or \{\Ar , \Ite{itny}{\True}, \Ex_1^{\Inf}\True\} \Or \TNArr ) \\[\NL]
                   & \And \TNBool) \\[\NL]
   & \Or \{\Ar , \Ite{itny}{\True}, \Ex_1^{\Inf}\True\} \Or \TNArr \\[\NL]
\Def{not\_y}{}&(\{\Ar , \Ite{itny}{\True}, \Ex_1^{\Inf}\True\} \Or \TNArr) \\[\NL]
\Def{itny}{}&\TNNum \And ((\{\Ar , \Ite{\Ref{itx}}{\True}\}\And \{\Ar , \Ite{\Ref{ity}}{\True}\}) \\[\NL]
      & \Or \TBo) \\[\NL]
\end{array}$$

After expansion, variables are only found in guarded positions and their definition is hence a bolean combination
of typed groups, hence we are now ready to transform the schema into a Disjunctive
Normal Form.

\comment{At the end of this process, the schema will respect the following normalized 
grammar.
$$
\begin{array}{llll}
S & ::= &\ \Type(T) 
              \M \Enu(J_1,\ldots,J_n) \\[\NL]
&& \M \Len_{i}^{j} \M \Pat(r) \M \NotPat(r) \\[\NL] 
&& \M \Bet_{m}^{M}  \M \XBet_{m}^{M}  \M \Mof(n) \M \NotMof(n) \\[\NL]
&&  \M \Props(r_1 : \Ref{x}_1,\ldots,r_n : \Ref{x}_n; \Ref{x}) \M 
      \Pro_{i}^{j}  \\[\NL]
      && \M \Req(k_1,\ldots,k_n) 
  \M \PReq(r_1,\ldots,r_n) \M \APReq(r_1,\ldots,r_n) \M  
        \\[\NL]
&&  \M \Ite{\Ref{x}_1 \itdots \Ref{x}_n}{\Ref{x}_{n+1}} \M \Ex_{i}^{j}\Ref{x} \M  \Uni \M \NotUni  \\[\NL]
&& \M \ADef{x_1}{S_1} , \ldots, \Def{x_n}{S_n} \M \Ref{x}  \M S_1 \And S_2  \M \Not S \\[\NL]
&&   \M \True  \M  \False \M  S_1 \Or S_2   \M  \XOr(S_1,\ldots,S_n)   \M S_1 \Implies S_2 \M  ( S_1 \Implies S_2 \ | \ S_3 )   \M \{ S_1, \ldots, S_n \} \\[\NL] 
&& \M S_1 \And S_2 \\[\NL]
T & ::= & \TNull \M \TBool \M \Str  \M \TNum \M \Int  \M
              \Arr \M \Obj \\[\NL] 
r & ::= &  \text{\em {\jsonsch} regular expression}  \\[\NL] 
J & ::= & \text{\em JSON expression}
\end{array}
$$}

\comment{
LIST OF OPERATORS

\section{List of base operators}
$$
\begin{array}{llll}

\Pat(r), \Bet_{m}^{M} , \Mof(n) , \IBT(b) \\
\key{r} :: S, \Pro_{i}^{j} ,  
\Ite{S_1 \itdots S_n}{S_{n+1}}, \Ex_{i}^{j}S, \Uni,\\
 S_1 \And S_2  , \Not S \\
\ADef{x_0} {S_0} , \ldots, \Def{x_n}{S_n}, \Ref{x}_i 
\end{array}
$$

Full list
$$
\begin{array}{llll}
\Type(T) ,  \Con(J)  ,  \Enu(J_1,\ldots,J_n), 
\Len_{i}^{j} , \Pat(r), \NotPat(r) , \Bet_{m}^{M} ,  \XBet_{m}^{M} , \Mof(n) ,
\NotMof(n), \\
\key{k} : S, \key{r} :: S, \AddP([k_1,\ldots,k_n],[r_1,\ldots,r_m]) :: S,
\Pro_{i}^{j} , \Req(k_1,\ldots,k_n),  \Nam(S), 
\PReq(r_1,\ldots,r_n),  \ExNam(S) ,\\
\Ite{S_1 \itdots S_n}{S_{n+1}}, \Ex_{i}^{j}S, \Uni,  \NotUni,\\
S_1 \And S_2  , \Not S, \True  , 
\False, S_1 \Or S_2  ,\XOr(S_1,\ldots,S_n)  ,S_1 \Implies S_2, ( S_1 \Implies S_2 \ | \ S_3 ),
\{ S_1, \ldots, S_n \},\\
\ADef{x_0}{S_0} , \ldots, \Def{x_n}{S_n}, \Ref{x}_i,
\end{array}
$$
}

\subsection{Transformation in Disjunctive Normal Form}\label{sec:DNF}

To reach a Disjunctive Normal Form (DNF), we repeatedly 
apply the following rule, and we apply and-merging and basic boolean reductions to any
new conjunction that is generated.
\setlength{\NL}{0.8ex}
$$
\begin{array}{llllllll}
\multicolumn{3}{l}{(S_{1,1}\Or\ldots\Or S_{1,n_1})\And
\ldots\And 
(S_{m,1}\Or\ldots\Or S_{m,n_m})}   \\
&=& \BigOr_{1\leq i_1 \leq n_1, \ldots, 1 \leq i_m \leq n_m}(S_{1,i_1}\And\ldots\And S_{m,i_m})
\\[\NL]
\end{array}
$$

\hide{
\begin{remark}
If we kept $\XOrVar$ and $\NotXOr$, we would have these three rules.
$$
\begin{array}{llllllll}
\multicolumn{3}{l}{(S_{1,1}\Or\ldots\Or S_{1,n_1})\And
\ldots\And 
(S_{m,1}\Or\ldots\Or S_{m,n_m})}   \\
&=& \BigOr_{1\leq i_1 \leq n_1, \ldots, 1 \leq i_m \leq n_m}(S_{1,i_1}\And\ldots\And S_{m,i_m})
\\[\NL]
\XOrVar(S_1,\ldots,S_n;S'_1,\ldots,S'_n) &=& 
\BigOr_{1 \leq i \leq n} ( S_i \And \BigAnd_{1 \leq j \leq n, j \neq i} S'_j ) \\[\NL]
\NotXOr(S_1,\ldots,S_n;S'_1,\ldots,S'_n) &=&
(S_1 \And \ldots \And S_n) \Or \BigOr_{1 \leq i < j \leq n} (S'_i \And S'_j) \\[\NL]
\end{array}
$$
\end{remark}
}

For example, we can apply DNF to the last definition of the previous example,
followed by and-merging and then by an extended-and-merging.
$$\begin{array}{rllll}
\multicolumn{3}{l}{\TNNum \And ((\{\Ar , \Ite{\Ref{itx}}{\True}\}\And \{\Ar , \Ite{\Ref{ity}}{\True}\}) \Or \TBo)} \\[\NL]
\key{DNF:} & = (\TNNum \And \{\Ar , \Ite{\Ref{itx}}{\True}\}\And \{\Ar , \Ite{\Ref{ity}}{\True}\})\\
& \qquad
\Or
(\TNNum \And \TBo) \\[\NL]
\key{AndM:} & = (\TNNum \And \{\Ar , \Ite{\Ref{itx} \And \Ref{ity}}{\True}\})
\\
& \qquad
\Or (\TNNum \And \TBo) \\[\NL]
\key{AndM:} & = \{\Ar , \Ite{\Ref{itx} \And \Ref{ity}}{\True}\} \Or \TBo \\[\NL]
\end{array}$$

\hide{
\begin{remark}
In the last step we simplify 
$\TNNum \And \{\Ar , \Ite{\Ref{itx} \And \Ref{ity}}{\True}\}$
as
$\{\Ar , \Ite{\Ref{itx} \And \Ref{ity}}{\True}\}$

If we consider $\TNNum$ as a unit, this is an application of a new rule of and-merging,
that combines a disjunction of type declarations $D$ with a typed group $TG$:
$$D, \key{TG} \To \key{TG} \quad\text{ if }\quad \TypeOf(\key{TG}) \in D$$

Literally, it is a complex step of DNF reduction followed by and-merging:
$$\begin{array}{llll}
& \TNNum \And \{\Ar , \Ite{\Ref{itx} \And \Ref{ity}}{\True}\} \\[\NL]
= &(\TNl\Or \TBo \Or \TSt \Or \TOb \Or \TAr) \And \{\Ar , \Ite{\Ref{itx} \And \Ref{ity}}{\True}\} \\[\NL]
= &(\TNl \And \{\Ar , \Ite{\Ref{itx} \And \Ref{ity}}{\True}\} ) \Or
(\TBo \And \{\Ar , \Ite{\Ref{itx} \And \Ref{ity}}{\True}\} ) \Or\\[\NL]
&(\TSt \And \{\Ar , \Ite{\Ref{itx} \And \Ref{ity}}{\True}\} ) \Or
(\TOb \And \{\Ar , \Ite{\Ref{itx} \And \Ref{ity}}{\True}\} ) \Or \\[\NL]
&(\TAr \And \{\Ar , \Ite{\Ref{itx} \And \Ref{ity}}{\True}\} ) \\[\NL]
=&\False \Or \False \Or \False \Or \False \Or  \{\Ar , \Ite{\Ref{itx} \And \Ref{ity}}{\True}\}   \\[\NL]
=&\{\Ar , \Ite{\Ref{itx} \And \Ref{ity}}{\True}\}   \\[\NL]
\end{array}$$
\end{remark}
}

Canonicalization ensures that all groups are typed groups, hence that they 
contain a $\Type(T)$ assertion.
Guarded expressions separation and variable expansion ensure that all guarded schemas 
are variables
(separation invariant)
and that no boolean expressions involve variables (expansion invariant).
Reduction in DNF preserves these invariants.
Unfortunately, the and-merging phase inserts conjunctions of variables in guarded positions, which breaks the separation and expansion invariants. This is not a problem, since object and array preparation have the same effect,
hence, after the next step, we have to go back to variable normalization, canonicalization,
preparation, until convergence.

Observe that all the arguments of any disjunction left
after canonicalization, variable normalization and reduction to DNF 
are typed groups.
Now we need to prepare these typed groups for the witness generation phase.

\subsection{The structure of the preparation phase} \label{sec:prep}

Before starting witness generation, we must bring the object groups (the typed
groups with type $\Obj$) and the array groups in a simplified form where the interactions
between the different components are explicit.
This is described in detail in the next two sections.

\subsubsection{Object group preparation} \label{sec:objprep}

Object type preparation has similarities with and-merging, but is different. And-merging is
an optimization, which performs some easy and optional rewritings that reduce the size of the
expression. Type preparation is mandatory, since it provides the type with the completeness 
and no-overlapping
invariants that are needed for witness generation. The similarity between the two phases
derives from the fact that they are based on similar equivalences, since they
are both applying semantics-preserving transformations to a conjunction of typed assertions. 

\comment{
This is the pseudo-code for the algorithm we are going to describe.
\begin{tabbing}
aa\=aa\=aa\=aa\=aa\=aa\=\kill
TranslateIntoCore \\
CompleteConstrainingPart; \\
While (exist p1,p2 in CP with overlap(p1,p2) \\
    ApplyRuleOne(CP,p1,p2); \\
For (pattReq(rj,Spj) in RP, pi:Si in CP) applyRule2; \\
while (exist r1,r2 in RP with overlap(p1,p2) and overlap(s1,s2) \\
    ApplyRuleThree(RPr1,r2); \\
NormalizeAllVariables; \\
\end{tabbing}}

For this phase and the next one, we introduce 
a new operator $$\OPReq(r_1 : S_1,\ldots,r_{n} : S_{n})$$
that represents the disjunction 
$$\PReq(r_1 : S_1)\Or\ldots\Or\PReq(r_{n} : S_{n})$$
and describes $n$ possible ways of satisfying a single $\PReq$ constraints.

The aim of object preparation is to bring object groups into a form where
one can easily enumerate all possible ways of satisfying all different assertions, by 
making all the interactions between different assertions explicit.

More precisely, we rewrite each object group into a \emph{constraining set} 
$$p_1 : \Ref{x}_1,\ldots,p_n : \Ref{x}_n$$ and a \emph{requiring set}
$$\begin{array}{lll}
\OPReq(r^1_1 : \Ref{y}^1_1,\ldots,r^1_{n_1} : \Ref{y}^1_{n_1}),\ldots,\\
\OPReq(r^m_1 : \Ref{y}^m_1,\ldots,r^m_{n_m} : \Ref{y}^m_{n_m})
\end{array}$$
which satisfy the following four properties.
Hereafter we say that two patterns $r_1$ and $r_2$ have a \emph{trivial intersection}
when either $r_1 \AndP r_2 = \FalseP$ or $r_1 = r_2$, 
so that they are either disjoint or equivalent.
\begin{enumerate}
\item Constraint partition: the patterns $p_i$ in the constraining part
   are mutually disjoint and cover all names.
\item Constraint internalization in the requiring part: for any pair $r^l_k : \Ref{y}^l_k$ in the 
    requiring part, and for each pair $p_i : \Ref{x}_i$ in the constraining part
    such that $r^l_k \AndP p_i \neq \FalseP$,
    we have that $\Ref{y}^l_k\subseteq \Ref{x}_i$. In this way, when $\Ref{y}^l_k$ is satified,
    all the constraints $p_i : \Ref{x}_i$ that apply to some name that matches
    $r^l_k$ are guaranteed to be satisfied, hence thay are \emph{internalized}
    in the assertion $\Ref{y}^l_k$.
\item Requirements internal splitting: for any two distinct pairs $r^i_j : \Ref{y}^i_j$ and
     $r^i_k : \Ref{y}^i_k$ inside the same $\OPReq$, they are \emph{split}, which means that:
     \begin{enumerate}
		\item  the two patterns have a \emph{trivial intersection}, and
        \item  either the pairs are pattern-disjoint,
     that is $r^i_j \AndP r^i_k = \FalseP$, or they are schema-disjoint, that is
         $\Ref{y}^i_j \And \Ref{y}^i_k = \False$.
 	\end{enumerate}
\item Requirements external splitting: any two distinct pairs $r^i_j : \Ref{y}^i_j$ and
     $r^l_k : \Ref{y}^l_k$  found in two distinct $\OPReq$ are either split, as in the previous
      definition, or equal in both components. 
\end{enumerate}

These invariants depend on inclusion assertion, as in $\Ref{y}^l_k\subseteq \Ref{x}_i$,
or disjointness, as in $\Ref{y}^i_j \And \Ref{y}^i_k = \False$. During preparation, we are not going to \emph{check} whether two assertions are included or disjoint, since this would be as hard as checking
satisfiability. We are going to \emph{build} assertions that satisfy this constraints, by adding a 
factor $\_\And \Ref{x}$ when we need a sub-assertion of $\Ref{x}$, or a factor $\_\And \Not(\Ref{x})$
when we need disjunction from $\Ref{x}$.

\renewcommand{\PP}[1]{\NN\,{#1}}
\renewcommand{\PPP}[1]{\NN{#1}\$}
\renewcommand{\CProps}[1]{#1}

As an example, consider the following group.
We use here JSON regular expressions, where $\hat{}\,$ matches the beginning of 
a string, $[\NN abc]$ matches any one character different from $a$, $b$ and $b$,
a dot $.$ matches any character, $\$$ matches the end of the string, so that 
``$\PP{a[\NN b].}$'' matches $acccccc$ and $acc$ but does not match $ac$, because the
dot after the ``$\PP{a[\NN b]}$'' requires a third letter (please look carefully for the dots in the patterns). This is the group.
$$\begin{array}{llll}
\{ \Ob, \CProps{\PP{a} : \Ref{x1}}, \CProps{\PP{.b} : \Ref{x2}},
            \PReq(\PP{.d} :\True), \PReq(\PP{a} :\Ref{x3} ) 
            \}
\end{array}$$
Object preparation will first rewrite it as follows, and it will then
create new variables to separate and expand all conjunctions such as 
$\Ref{x1} \And \Ref{x2}$.
The variable $\NR{x3}$ is the variable whose body is the negation of that of $\Ref{x3}$.
The step-by-step process that produces this expansion is described in \cite{long}, but we show here
the final result.
$$\begin{array}{llll}
\CProps{\PP{a} : \Ref{x1}}, \CProps{\PP{.b} : \Ref{x2}}
&\To&
\PP{a[\NN b]} : \Ref{x1} , \PP{ab} : \Ref{x1} \And \Ref{x2}, \\
&&  \PP{[\NN a]b} :  \Ref{x2}, 
 \PP{[\NN a][\NN b]} : \True
\\[\NL]
\PReq( \PP{.d}: \True) &\To&
         \OPReq( \PP{ad} : \Ref{x1} \And \Ref{x3} , \\
         &&\qquad\qquad\quad
         \PP{ad}  : \Ref{x1} \And \NR{x3}, 
         \PP{[\NN a]d} : \True)\\[\NL]
\PReq( \PP{a}: \Ref{x3}) &\To&
          \OPReq( \PP{ad}: \Ref{x1} \And \Ref{x3}, 
                      \PP{a[\NN bd]} : \Ref{x1} \And \Ref{x3}, \\[\NL]
                     &&\qquad\qquad\quad \PP{ab}  : \Ref{x1} \And \Ref{x2} \And \Ref{x3} ),  
\end{array}$$
In the constraining part, the set $\{\PP{a}, \PP{.b}\}$ has been divided into three disjoint
parts $\{ \PP{a[\NN b]} , \PP{ab}, \PP{[\NN a]b} \}$ by separating the intersection
$\PP{ab}$ from the two original patterns,
and the set is completed with $\PP{[\NN a][\NN b]}  : \True$.
The first request $\PReq(\PP{ .d} : \True)$ is split into three different cases.
The first $\PP{ad} : \Ref{x1} \And \Ref{x3}$ is in common with the other
$\OPReq$, while the case $\PP{ad}  : \Ref{x1} \And \NR{x3}$
is internally and externally split thanks to the $\NR{x3}$ factor
in the schema, and $\PP{[\NN a]d} $ is  
pattern-disjoint thanks to the initial $[\NN a]$. 
You can also observe that $\PP{ad}  : \Ref{x1} \And \Ref{x3}$ internalizes the requirement
$\PP{a} : \Ref{x1}$, the same holds for $\PP{ad}  : \Ref{x1} \And \NR{x3}$,
while $\PP{[\NN a]d}$ only matches the trivial requirement, hence maintains its 
$\True$ schema.
The second $\PReq$ is split into three cases as well, in order to bring into view the intersection
with the first $\PReq$, and in order to internalize the constraints of the constraining part.

This splitting effort is needed in order to be able to enumerate and try all the possible ways of 
satisfying a set of requests. For example, in this case the two $\OPReq$ requests share the first 
component $\PP{ad}: \Ref{x1} \And \Ref{x3}$, and contain two more components each, all of them
mutually incompatible, hence having a structure $\OPReq(a,b1,b2),\OPReq(a,c1,c2)$. Hence, we know that 
there are exactly 5 ways of satisfying both: either by generaing a single member that satisfies $a$, or by generating
two members that satisfy, respectively, $(b1,c1)$, $(b1,c2)$, $(b2,c1)$, $(b2,c2)$, and our witness generation
algorithm will try to pursue all, and only, these five approaches.

Also array groups need preparation, still because several assertions inside an array group may overlap. For space reason we omit this part in this paper (details are in the full version \cite{long}).

\subsection{Recursive witness generation}\label{sec:recgen}

\renewcommand{\Next}{\kw{next}}
\renewcommand{\EOV}{\kw{EOV}}
\renewcommand{\DOV}{\kw{DOV}}

We  illustrate the algorithm by means of  an example, by focusing on object groups (other cases are dealt with in the full paper \cite{long}). Consider the following
set of equations, which is not complete since we removed all those that are not 
reachable from the root.
$$\begin{array}{llll}
\ADef{\Ref{x}}{}&\!\!\{ \Ob, \OPReq(\PP{a} : \Ref{l}, \PP{b} : \Ref{y}) \}\\[\NL]
\Def{\Ref{y}}{}&\!\!\{ \Ob, \OPReq(\PP{a} : \Ref{z}, \PP{b} : \Ref{k}), \OPReq(\PP{c} : \Ref{m}) \}\\[\NL]
\Def{\Ref{k}}{}&\!\!\{ \Ob, \OPReq(\PP{a} : \Ref{l}) \}\\[\NL]
\Def{\Ref{l}}{}&\!\!\{ \Ob, \OPReq(\PP{a} : \Ref{x}), \Pro_0^0 \}\\[\NL]
\Def{\Ref{z}}{}&\!\!\{ \Ob, \OPReq(\PP{a} : \Ref{x}) \} \Or \Nl\\[\NL]	
\Def{\Ref{m}}{}&\!\!\TNu\\[\NL]		
\end{array}$$

The algorithm proceeds by passes. Each pass begins with a state where each variable
is either \Pop\ with a witness $J$ ($P(J)$), \Emp\ ($\False$) if we proved that it has no witnesses, or
\Open\ (?), otherwise. At the beginning each variable is \Open:
$$
P_0 : x = ?,
y = ?,
k = ?,
l = ?,
z = ?, 
m = ?
$$
At each pass, we evaluate the body of each variable using the state of the previous pass.
If nothing changes, we stop. Otherwise, we continue until a witness is found for the root
variable.
Here, at pass 1, we are able to prove that \Ref{l} is empty, and we can provide a witness for
$\Ref{z}$ and $\Ref{m}$.
$$
P_1 : x = ?,
y = ?,
k = ?,
l = \False,
z = P(\xnull), 
m = P(3)
$$
At pass 2 we use the knowledge of pass 1 to 
fix both $\Ref{y}$ and $\Ref{k}$.
$$
P_2 : x = ?,
y = P(\{ a : \xnull, c : 3 \}), 
k = \False,
l = \False,
z = P(\xnull), 
m = P(3)
$$
And finally, we converge at pass 3.
$$\begin{array}{lll}
P_3 : &x = P(\{ b : \{ a : \xnull, c : 3 \} \}), 
y = P(\{ a : \xnull, c : 3 \}), \\
&k = \False,
l = \False,
\ldots
\end{array}
$$

As a negative example, consider the following system
$$\begin{array}{llll}
\ADef{\Ref{x}}{}&\!\!\{ \Ob, \OPReq(\PP{a} : \Ref{y}) \}\\[\NL]
\Def{\Ref{y}}{}&\!\!\{ \Ob, \OPReq(\PP{a} : \Ref{z}) , \OPReq(\PP{b} : \Ref{x}) \}\\[\NL]
\Def{\Ref{z}}{}&\!\!\{ \Ob, \OPReq(\PP{a} : \Ref{y}) \} \Or \TNu\\[\NL]
\end{array}$$

We report here the trace of a run of the algorithm.
$$\begin{array}{llll}
P_0 : x = ?, y = ?, z = ?\\[\NL]
P_1 : x = ?, y = ?, z = P(3)\\[\NL]
P_2 :  x = ?, y = ?, z = P(3)\\[\NL]
\kw{stop}
\end{array}$$
Here, the algorithm reached a fix point without a value for $\Ref{x}$ and $\Ref{k}$, which are therefore
empty.

Hence, the algorithm is defined as follows. We first mark all variables that are actually
reachable from the root, and delete the others. We associate a state of \Open\ to each
variable.
We compute a new state for each \Open\ variable
on the basis of the current state.
If the new state is equal to the previous state, the algorithm returns ``no witness''.
If the new state has a witness for the root variable, this witness is returned.
If the state changed but the root variable is still \Open, we execute a new pass.

The computation of the new state proceeds as follows. 
The definition of each variable is a disjunction of typed groups. The variable
asks a witness to each of them. If one group provides a witness, this is a witness for the variable.
If one group answers $\False$, it is removed and, if all are removed, the variable answers $\False$.
If all groups answer $?$, then the variable is still \Open.
For the typed groups, each runs an algorithm that depends on its type, which is described
below. For the base groups, the answer is either $P(J)$ or $\False$ during the first
pass, and will not change. For the object and array groups the answer will depend on
the current state of the variables that appear in them.

Termination of recursive witness generation can be proved using a classical minimal-fixpoint
argument, as follows.

Witness generation returns three possible results, \Emp, \Open, and \Pop, and every 
variable, at the end of each pass, is in one of those three states. If we 
order these states as $\Open < \Pop$ and $\Open < \Emp$, we observe that witness 
generation for object
and array types is a monotone function on the state. In greater detail, the 
\Open, \Emp, or \Pop\ result of witness generation is uniquely determined by the state
 of all variables, and, whenever one variable increases its state, the result of witness generation
 either remains equal or increases. As a consequence, the trace of any run, defined as the 
 sequence of tuples that associate each variable with its state, can only increase or remain 
 immobile at every step hence, having a finite number of distinct values, the trace is 
 guaranteed to converge to a fixpoint.

\subsection{Witness Generation from Typed Groups}

We have finally to specify how each typed group will generate its witnesses starting from the witnesses 
associated to the different variables. The treatment for these cases is dealt with in the full version \cite{long}.


\section{Related Work}\label{sec:relworks}
We are not aware of any formal algebra for JSON Schema.
The first effort to formalize the semantics of JSON Schema as by Pezoa et al. in  \cite{DBLP:conf/www/PezoaRSUV16} whose goal was to lay the foundations of the JSON schema proposal by studying its expressive power and the complexity of the validation problem. Along the lines of this work, Bouhris et al.  \cite{DBLP:conf/pods/BourhisRSV17} characterized the expressivity of the JSON Schema language and investigated the complexity of the satisfiability problem which turns out to be \emph{2EXPTIME} in the general case and \emph{EXPSPACE} when disallowing \emph{uniqueItems}. 
None of the above works study the problem of generating an instance of a JSON Schema.
The only attempt to solve this problem was investigated by Earle et. al \cite{benac2014jsongen} in the context of testing REST API calls but the presented solution, which is based on translating JSON Schema definitions into an Erlang expression, is not formally defined and restricted to atomic values, objects and to some form of boolean expressions. 

From the point of view of schema normalization, the closest work to ours is the one in \cite{habib2019type} which studies schema inclusion for JSON Schema. 
To cope with the high expressivity of the JSON Schema language, a pre-requisite step is needed to rewrite the schemas into a \emph{Disjunctive Normal Form} which has some similarities with the preparation phase of our work.
However, compared to our work, the schema normalization in \cite{habib2019type} lacks the ability of eliminating negation for all kinds constraints, does not deal with recursive definitions and is not able to decide schema satisfiability which is captured by the \emph{inhabited( )} predicate whose specification is only informally discussed.
This has been confirmed in practice by experimenting the tool developed in \cite{habib2019type} for parsing real world schemas described in \cite{long}:  the tool raised an issue for 21,859 out of 23,480 input schemas. The dominating error is related to  constructs not being supported, but many other errors due to the inability to parse recursive schemas or to navigate references  are present.




\section{Conclusions}\label{sec:concl}

\jsonsch\/ is an evolving standard for the description of families of JSON documents, and is widely used in data-centric applications. Despite the recent interest in the research community related to this schema language, crucial problems like schema equivalence/inclusion and consistency have either been partially dealt with or not explored at all. 
In this work we present our approach in order to solve these problems, based on our algebraic specification of JSON Schema. We are currently finalising a Java implementation of the presented algorithm, and studying optimisation techniques, by analysing a large repository of JSON Schemas allowing us for determining how often mechanisms that are critical for execution times are used. We are also investigating witness generation techniques able to generate several instances meant  be used for testing queries and programs manipulating valid JSON data. 

\section*{Acknowledgements}
The research has been partially supported by the MIUR project PRIN 2017FTXR7S ``IT-MaTTerS'' (Methods and Tools for Trustworthy Smart Systems)  and by the Deutsche Forschungsgemeinschaft (DFG, German Research Foundation), grant \#385808805.

%



\appendix

\end{document}